\documentclass[fleqn,usenatbib]{mnras}

\usepackage{newtxtext,newtxmath}

\usepackage[T1]{fontenc}

\DeclareRobustCommand{\VAN}[3]{#2}
\let\VANthebibliography\thebibliography
\def\thebibliography{\DeclareRobustCommand{\VAN}[3]{##3}\VANthebibliography}

\usepackage{graphicx}	
\usepackage{amsmath}	
\usepackage{amssymb}	

\usepackage{bm} 

\graphicspath{{Figures/}}

\newcommand{\alm}{a_{\ell m}}
\newcommand{\Cl}{C_\ell}
\newcommand{\lmax}{\ell_\text{max}}
\newcommand{\lmin}{\ell_\text{min}}
\newcommand{\leff}{\ell_\text{eff}}

\title[Accuracy of Gaussian weak lensing likelihood]
{Sufficiency of a Gaussian power spectrum likelihood for accurate cosmology from upcoming weak lensing surveys}

\author[R. E. Upham, M. L. Brown and L. Whittaker]{
Robin E. Upham,$^{1}$\thanks{E-mail: robin.upham@manchester.ac.uk}
Michael L. Brown$^{1}$
and Lee Whittaker$^{\,2, 1}$
\\
$^{1}$Jodrell Bank Centre for Astrophysics, Department of Physics and Astronomy, University of Manchester, Oxford Road, Manchester M13 9PL, UK\\
$^{2}$Department of Physics and Astronomy, University College London, Gower Street, London WC1E 6BT, UK
}

\date{Accepted 2021 February 18. Received 2021 February 18; in original form 2020 December 11}

\pubyear{2021}

\begin{document}
\label{firstpage}
\pagerange{\pageref{firstpage}--\pageref{lastpage}}
\maketitle

\begin{abstract}
We investigate whether a Gaussian likelihood is sufficient to obtain accurate parameter constraints from a \textit{Euclid}-like combined tomographic power spectrum analysis of weak lensing, galaxy clustering and their cross-correlation.
Testing its performance on the full sky against the Wishart distribution, which is the exact likelihood under the assumption of Gaussian fields, we find that the Gaussian likelihood returns accurate parameter constraints. 
This accuracy is robust to the choices made in the likelihood analysis, including the choice of fiducial cosmology, the range of scales included, and the random noise level.
We extend our results to the cut sky by evaluating the additional non-Gaussianity of the joint cut-sky likelihood in both its marginal distributions and dependence structure. We find that the cut-sky likelihood is more non-Gaussian than the full-sky likelihood, but at a level insufficient to introduce significant inaccuracy into parameter constraints obtained using the Gaussian likelihood.
Our results should not be affected by the assumption of Gaussian fields, as this approximation only becomes inaccurate on small scales, which in turn corresponds to the limit in which any non-Gaussianity of the likelihood becomes negligible.
We nevertheless compare against N-body weak lensing simulations and find no evidence of significant additional non-Gaussianity in the likelihood. 
Our results indicate that a Gaussian likelihood will be sufficient for robust parameter constraints with power spectra from Stage IV weak lensing surveys.
\end{abstract}




\section{Introduction}

Analysis of weak gravitational lensing of distant galaxies by large scale structure is among the most promising methods of constraining theories of dark energy in the near future. Upcoming surveys such as those with \textit{Euclid}\footnote{\href{https://www.euclid-ec.org}{https://www.euclid-ec.org}} \citep{Laureijs2011}, the Vera C. Rubin Observatory (LSST)\footnote{\href{https://www.lsst.org}{https://www.lsst.org}} \citep{Ivezic2019} and the Square Kilometre Array (SKA)\footnote{\href{https://www.skatelescope.org}{https://www.skatelescope.org}}  \citep{SKA2018} will estimate the shapes and redshifts of ${\sim}10^9$ galaxies, an order of magnitude increase on the current generation comprising the Dark Energy Survey (DES)\footnote{\href{https://www.darkenergysurvey.org}{https://www.darkenergysurvey.org}} \citep{DES2005}, Kilo-Degree Survey (KiDS)\footnote{\href{http://kids.strw.leidenuniv.nl}{http://kids.strw.leidenuniv.nl}} \citep{DeJong2013} and Hyper Suprime-Cam (HSC)\footnote{\href{https://hsc.mtk.nao.ac.jp}{https://hsc.mtk.nao.ac.jp}} \citep{Miyazaki2012}. However, the unprecedented statistical precision offered by such experiments requires equally unprecedented control of sources of systematic error in order to obtain reliable results. One of the many such sources is the choice of likelihood function, currently routinely assumed to be Gaussian \citep[e.g.][]{Troxel2018, Hikage2019, Joachimi2020}.

However, the true likelihood of weak lensing two-point statistics is well known to be non-Gaussian. This has been studied in detail in distributions of simulated data \citep{Sellentin2018, Sellentin2018a, DiazRivero2020, Louca2020} and has motivated many derivations of non-Gaussian likelihoods, either approximate or exact under particular conditions \citep{Taruya2002, Sato2010, Sato2011, Hilbert2011, Keitel2011, Wilking2013, Sellentin2015, Wilking2015, Upham2019a, Manrique-Yus2020, DiazRivero2020}. 

The impact of wrongly assuming a Gaussian likelihood on cosmological parameter constraints has, however, rarely been investigated in detail. \cite{Lin2020} did so for the shear correlation function in an LSST-like experiment, and found that a Gaussian likelihood is sufficiently accurate for obtaining joint posterior constraints on $\Omega_\text{m}$ and $\sigma_8$, despite the small $100\,\text{deg}^2$ sky patch used in their tests. This result is in contrast to the earlier work in \cite{Hartlap2009}, which found that a Gaussian correlation function likelihood could lead to biased constraints in the same parameters. \cite{Taylor2019} tested the impact of assuming a Gaussian likelihood for the full-sky shear power spectrum on joint constraints of $\Omega_\text{m}$ and $S_8 = \sigma_8 (\Omega_\text{m} / 0.3)^{0.5}$ and found negligible difference in the posterior distribution compared to a likelihood-free approach.

In this paper we test the impact of assuming a Gaussian likelihood for a \textit{Euclid}-like joint tomographic "3$\times$2pt" power spectrum analysis of weak lensing shear, galaxy clustering and their cross-correlation, on posterior dark energy constraints. We begin with a full-sky setup in \autoref{Sec:fullsky}, before extending our results to a cut sky in \autoref{Sec:cut_sky} and to non-Gaussian fields in \autoref{Sec:nongauss_fields}. We discuss our conclusions in \autoref{Sec:conclusions}.

\section{Full-sky likelihood}
\label{Sec:fullsky}

\subsection{Background}

The observable fields we consider are weak lensing shear and galaxy number overdensity. For the majority of this work we will treat these fields, as observed, using Gaussian statistics. 
This is an approximation, but we have reason to believe it to be a good one for the purposes of this study, which we discuss in \autoref{Sec:nongauss_fields}. It is also a necessary starting point, since the only conditions under which the exact joint power spectrum likelihood is both known and tractable is for Gaussian fields on the full sky. Therefore, we will first obtain results for Gaussian fields. We will argue that these results hold for real observable fields in \autoref{Sec:nongauss_fields}, where we also analyse the distribution of power spectrum estimates from N-body simulations.

\subsubsection{Wishart distribution}

For correlated Gaussian fields observed on the full sky, the set of observed $\Cl$s follows a Wishart distribution, independently for each $\ell$ (see \citealt{Percival2006} for a derivation in the case of cosmic microwave background temperature and polarisation). This distribution can be parametrised using the degrees of freedom $\nu$ and $p \times p$ scale matrix $\mathbfss{V}$, in which case the probability distribution function (PDF) for random matrix $\mathbfss{X}$ is \begin{equation}
f_\mathcal{W} \left( \mathbfss{X} | \nu, \mathbfss{V} \right) = 
\frac{| \mathbfss{X} |^{(\nu - p - 1) / 2} 
\exp [ -\text{trace}(\mathbfss{V}^{-1} \mathbfss{X})/2 ]}
{2^{\nu p / 2} |{\mathbfss X}|^{\nu / 2} \Gamma_p(\nu / 2)},
\end{equation}
where $\Gamma_p$ is the multivariate gamma function. For an $N$-bin tomographic 3$\times$2pt analysis, we can write the set of observed $\Cl$s as a $2N \times 2N$ symmetric matrix, $\widehat{\mathbfss{C}}_\ell$:
\begin{equation}
\widehat{\mathbfss{C}}_\ell = 
\begingroup
\setlength\arraycolsep{3pt} 
\begin{pmatrix}
\widehat{C}_\ell^{n (1) n (1)} & \widehat{C}_\ell^{n (1) E (1)} & \cdots 
& \widehat{C}_\ell^{n (1) n (N)} & \widehat{C}_\ell^{n (1) E (N)} \\
\widehat{C}_\ell^{n (1) E (1)} & \widehat{C}_\ell^{E (1) E (1)} & \cdots &
\widehat{C}_\ell^{E (1) n (N)} & \widehat{C}_\ell^{E (1) E (N)} \\
\vdots & \vdots & \ddots & \vdots & \vdots \\
\widehat{C}_\ell^{n (1) n(N)} & \widehat{C}_\ell^{E (1) n(N)} & \cdots & 
\widehat{C}_\ell^{n (N) n (N)} & \widehat{C}_\ell^{n (N) E (N)} \\
\widehat{C}_\ell^{n (1) E(N)} & \widehat{C}_\ell^{E (1) E(N)} & \cdots & \widehat{C}_\ell^{n (N) E (N)} & \widehat{C}_\ell^{E (N) E (N)} \\
\end{pmatrix},
\endgroup
\label{Eqn:obs_cl_matrix}
\end{equation}
where $n$ represents the number overdensity field and $E$ the shear $E$-mode, and $\widehat{C}_\ell^{X (i) Y (j)}$ is the observed cross-power between redshift bins $i$ and $j$. For Gaussian fields, $\widehat{\mathbfss{C}}_\ell$ follows a Wishart distribution with parameters
\begin{equation}
\widehat{\mathbfss{C}}_\ell \sim 
\mathcal{W} \left( \nu = 2 \ell + 1, 
\mathbfss{V} = \frac{\mathbfss{C}_\ell}{2 \ell + 1} \right),
\label{Eqn:wishart}
\end{equation}
where $\mathbfss{C}_\ell$ is the symmetric positive definite matrix of underlying $C_\ell$s analogous to $\widehat{\mathbfss{C}}_\ell$,
\begin{equation}
\mathbfss{C}_\ell = 
\begingroup
\setlength\arraycolsep{3pt} 
\begin{pmatrix}
C_\ell^{n (1) n (1)} & C_\ell^{n (1) E (1)} & \cdots 
& C_\ell^{n (1) n (N)} & C_\ell^{n (1) E (N)} \\
C_\ell^{n (1) E (1)} & C_\ell^{E (1) E (1)} & \cdots &
C_\ell^{E (1) n (N)} & C_\ell^{E (1) E (N)} \\
\vdots & \vdots & \ddots & \vdots & \vdots \\
C_\ell^{n (1) n(N)} & C_\ell^{E (1) n(N)} & \cdots & 
C_\ell^{n (N) n (N)} & C_\ell^{n (N) E (N)} \\
C_\ell^{n (1) E(N)} & C_\ell^{E (1) E(N)} & \cdots & C_\ell^{n (N) E (N)} & C_\ell^{E (N) E (N)} \\
\end{pmatrix}.
\endgroup
\label{Eqn:theory_cl_matrix}
\end{equation}
The order of rows and columns in $\mathbfss{C}_\ell$ and $\widehat{\mathbfss{C}}_\ell$ is arbitrary, provided it is consistent between the two matrices. For simplicity we have ignored shape noise in \autoref{Eqn:obs_cl_matrix} and \autoref{Eqn:theory_cl_matrix}, but this may be included by replacing each $C_\ell$ in the diagonal with $C_\ell + N_\ell$, where $N_\ell$ is the corresponding noise power. We include noise in our \textit{Euclid}-like setup described in \autoref{Sec:fs_method}. This setup may also be trivially extended to include a shear $B$-mode.

It follows that the exact likelihood for a set of observed power spectra from correlated Gaussian fields on the full sky is a product of Wishart distributions, one for each $\ell$, each following \autoref{Eqn:wishart}.

\subsubsection{Gaussian distribution}

The multivariate Gaussian distribution, parametrised by mean vector $\bm{\mu}$ and covariance matrix $\bm{\Sigma}$, for length-$k$ random vector $\mathbfit{x}$ has PDF
\begin{equation}
f_\mathcal{N} \left( \mathbfit{x} | \bm{\mu}, \bm{\Sigma} \right)
= \left( 2 \pi \right)^{- k / 2}
| \bm{\Sigma} |^{-1/2}
\exp \left[ - \frac{1}{2} \left( \mathbfit{x} - \bm{\mu} \right)^\mathsf{T}
\bm{\Sigma}^{-1} \left( \mathbfit{x} - \bm{\mu} \right) \right].
\label{Eq:gauss_pdf}
\end{equation}
We may define a vector of observed $\Cl$s containing the unique elements of $\widehat{\mathbfss{C}}_\ell$. If $\widehat{\mathbfss{C}}_\ell$ obeys \autoref{Eqn:wishart}, then the expectation value of this vector will be the corresponding elements of $\mathbfss{C}_\ell$; i.e., the expectation value of any given observed $\widehat{C}_\ell$ is the corresponding underlying $C_\ell$. The covariance matrix of this vector has elements given by the well-known general expression for the covariance of full-sky $\Cl$ estimates,
\begin{equation}
\text{Cov} \left(
\widehat{C}_\ell^{\alpha \beta},
\widehat{C}_{\ell'}^{\gamma \varepsilon}
\right)
= \frac{\delta_{\ell \ell'}}{2 \ell + 1} \left(
C_\ell^{\alpha \gamma} C_\ell^{\beta \varepsilon}
+ C_\ell^{\alpha \varepsilon} C_\ell^{\beta \gamma} \right),
\end{equation}
where $\delta$ is the Kronecker delta. Therefore, we may approximate the exact distribution of full-sky power estimates (\autoref{Eqn:wishart}) with a Gaussian distribution having the same mean and covariance. 

It turns out that this approximation performs much better if the covariance is fixed at some fiducial cosmology, rather than being re-evaluated at each set of theory $\Cl$s being considered in a likelihood analysis. This is explored in some detail in \cite{Hamimeche2008} and \cite{Carron2013},
where it is also shown that allowing the covariance to vary as a function of cosmology violates the Cram\'er-Rao bound. This is also discussed in the methodology paper of the KiDS-1000 analysis \citep{Joachimi2020}. Therefore, this is the approximation that we test in this paper: when we refer to the ``Gaussian likelihood'', it is the version of \autoref{Eq:gauss_pdf} where $\bm{\Sigma}$ is fixed at some fiducial cosmology. We explore the impact of the choice of fiducial cosmology in \autoref{Sec:robustness}.

As we will discuss in more detail in \autoref{Sec:ma_marginals}, the marginal distributions of a Gaussian-distributed vector have zero skewness and excess kurtosis, which is not the case for the Wishart distribution. Since we are fixing the mean and variance of the Gaussian distribution to be equal to those of the Wishart distribution, the inaccuracy of the Gaussian likelihood approximation in describing the true marginal distributions will be largely captured by the skewness and excess kurtosis. However, for the Wishart distribution both of these quantities decrease as power laws in $2 \ell + 1$, and the behaviour of the cut-sky likelihood is similar. Therefore, the inaccuracy of the Gaussian likelihood will be most pronounced for low $\ell$, corresponding to large physical scales. We refer the reader to \autoref{Sec:ma_marginals} for more details.

\subsection{Full sky: Methodology}
\label{Sec:fs_method}

The tests in this section involve comparing exact posterior distributions, obtained with the Wishart likelihood, to approximate posterior distributions obtained with the Gaussian likelihood. We study the mean, maximum and standard deviation of single-parameter posteriors in \autoref{Sec:fs_sumstat} and the contours of two-dimensional posteriors in \autoref{Sec:fs_contours}. We calculate the posterior distribution of model parameters $\bm{\theta}$ from observed data $\mathbfit{d}$, $p \left( \bm{\theta} \,\middle|\, \mathbfit{d} \right)$, by evaluating Bayes' theorem,
\begin{equation}
p \left( \bm{\theta} \,\middle|\, \mathbfit{d} \right)
\propto \pi \left( \bm{\theta} \right)
f \left( \mathbfit{d} \,\middle|\, \bm{\theta} \right).
\end{equation}
The normalisation constant is formally given by the Bayesian evidence, but here we normalise manually assuming a uniform prior $\pi \left( \bm{\theta} \right)$, chosen to be sufficiently broad as to negligibly affect the posterior distribution. The remaining three ingredients are the model predictions, which are deterministic functions of $\bm{\theta}$, the (mock) observation $\mathbfit{d}$, and the likelihood function $f \left( \mathbfit{d} \,\middle|\, \bm{\theta} \right)$ which connects them. We describe each of these below.

\subsubsection{Theory}
\label{Sec:fs_method_theory}

We use regular grids of one, two or three cosmological parameters from ($w_0$, $w_a$, $\Omega_\text{m}$), with all other parameters held to a fixed value. We generated these grids using CosmoSIS\footnote{\href{https://bitbucket.org/joezuntz/cosmosis}{https://bitbucket.org/joezuntz/cosmosis}} \citep{Zuntz2015}. The pipeline consisted of the following CosmoSIS standard library modules:
\begin{enumerate}
\item CAMB version Jan15, to calculate the linear matter power spectrum \citep{Lewis2000, Howlett2012};
\item Halofit\_Takahasi version Camb-Nov-13, to compute the non-linear matter power spectrum (\citealt{Smith2003}, \citealt{Takahashi2012}; CosmoSIS module by A. Lewis \& S. Bird);
\item no\_bias version 1, to calculate the galaxy power spectrum with no galaxy bias -- this choice was made for simplicity, since galaxy bias is irrelevant to our tests;
\item gaussian\_window version 1, to calculate Gaussian redshift distributions -- we used 5 bins centred on $z =$ 0.65, 0.95, 1.25, 1.55, 1.85 each with $\sigma = 0.3$;
\item project\_2d version 1.0, to calculate projected galaxy and shear power spectra applying the Limber approximation -- the accuracy of the Limber approximation is also irrelevant for the purposes of our tests. We modified this module to output linearly spaced $\Cl$s for the full multipole range that we used ($2 \leq \ell \leq 2000$).
\end{enumerate}

\subsubsection{Mock observations}
\label{Sec:fs_method_obs}

To generate mock observations, we started by taking a set of output power spectra from the pipeline described above, with zero shear $B$-mode signal. We then added a noise contribution to each auto-power spectrum, $N_\ell$:
\begin{equation}
N_\ell^{n(i)n(i)} = \frac{1}{N_i}; 
\quad\quad
N_\ell^{E(i)E(i)} = 
N_\ell^{B(i)B(i)} = \frac{\sigma_\varepsilon^2}{N_i},
\end{equation}
where $N_i$ is the galaxy number density per redshift bin and $\sigma_\varepsilon$ is the intrinsic ellipticity dispersion per component. We use a \textit{Euclid}-like number density of $30 / \text{arcmin}^2$, split equally between the five redshift bins, and a value of $\sigma_\varepsilon = 0.3$.

This results in 120 input power spectra, of which 65 relate to shear $B$-mode so are zero or noise-only. From these we use the healpy Python implementation of the HEALPix\footnote{\href{https://healpix.sourceforge.io}{https://healpix.sourceforge.io}} software \citep{Gorski2005, Zonca2019} to generate 15 correlated maps, three per redshift bin, having the full set of 3$\times$2pt correlations, including the noise contribution and a proper spin-2 treatment of shear. The default resolution used in our tests is $n_\text{side} = 1024$ (corresponding to an angular scale of 3.4 arcmin) and $\lmax = 2000$; we note in the text whenever we depart from this. We use healpy to measure the 120 observed full-sky power spectra from these maps. 

We do not bin in $\ell$ to form bandpowers, since the exact likelihood in this case would no longer be a Wishart distribution. Instead it would follow a more complicated distribution, whose PDF could in principle be obtained either as a convolution of Wishart PDFs or from the general PDF of quadratic forms in Gaussian variables, analogous to the exact pseudo-$C_\ell$ likelihood derived in \cite{Upham2019a}. The feasibility of such an approach in practice is unclear, and is not the focus of this work. Furthermore, the distribution of individual $C_\ell$ estimates should be more non-Gaussian than the distribution of bandpowers, since each bandpower has more contributing modes. This implies that the results obtained here should be taken in this regard as a lower limit on the accuracy of the Gaussian likelihood.

\subsubsection{Likelihoods}

We exclude $B$-mode power spectra from the likelihood analysis, leaving 55 power spectra as input to the likelihoods. We implemented custom code in Python to evaluate each log-likelihood at every grid point. For the Wishart likelihood, we used the SciPy\footnote{\href{https://scipy.org}{https://scipy.org}} Wishart log-PDF function \citep{Virtanen2020}, which implements \autoref{Eqn:wishart}. For the Gaussian likelihood we used a custom implementation of the Gaussian log-PDF in \autoref{Eq:gauss_pdf} with precomputed inverse covariance, neglecting the determinant term since it is constant when the covariance is fixed. We exponentiate each log-likelihood and normalise each posterior distribution separately.

\subsection{Full sky: Summary statistics}
\label{Sec:fs_sumstat}

For the tests in this subsection, we generated 27\,000 mock observations following the steps outlined above (\autoref{Sec:fs_method_obs}), but with $\lmax = 100$ to keep computation time and data volume within reasonable limits. This isolates the part of the data vector for which the Gaussian likelihood should be expected to perform worst, since the likelihood is most non-Gaussian at low $\ell$ (see \autoref{Sec:cut_sky}). For each realisation, we ran a single-parameter likelihood analysis on $w_0$ with other parameters fixed. We now study the distributions of the maximum, mean and standard deviation of the resulting one-dimensional posterior distribution across all realisations.

\subsubsection{Posterior maximum}

\begin{figure*}
\centering
\includegraphics[width=.95\textwidth]{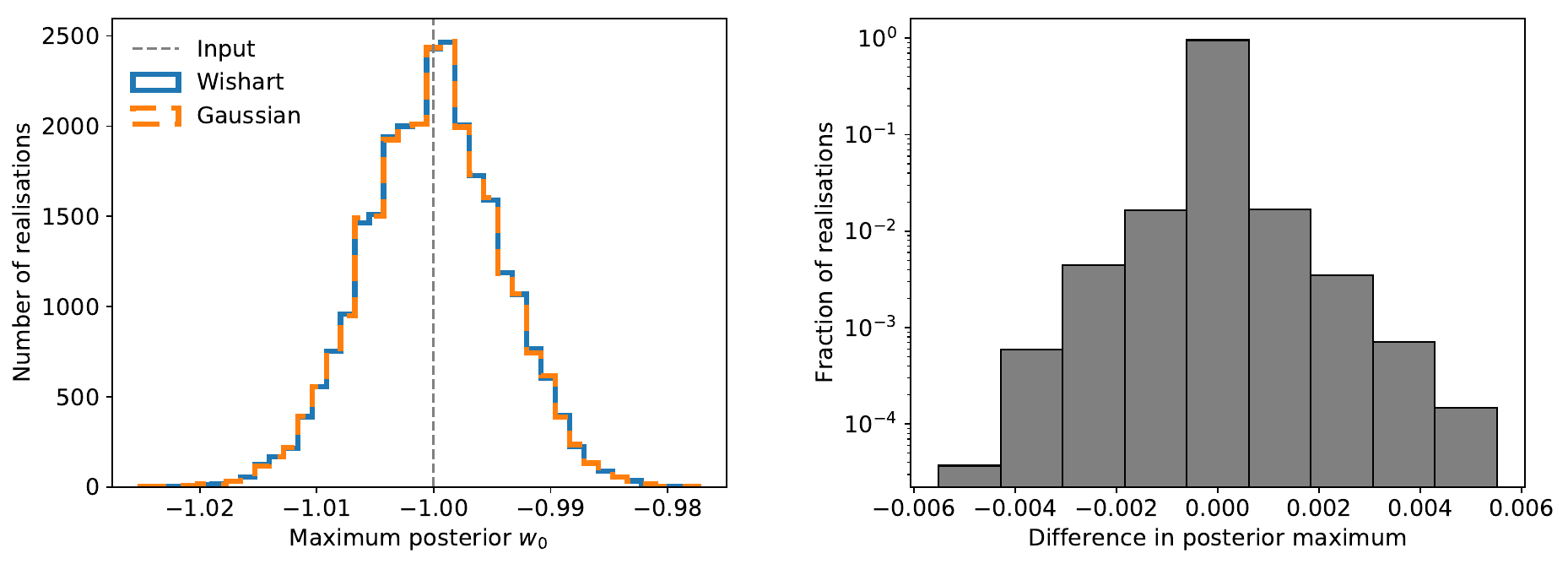}
\caption{\textit{Left:} Distribution of posterior maxima returned by the Gaussian likelihood compared to the true, Wishart likelihood. \textit{Right:} Per-realisation difference between the posterior maximum returned by the two likelihoods, with a positive difference representing a higher value of $w_0$ for the Gaussian likelihood.}
\label{Fig:postmax}
\end{figure*}

For a single spin-0 field, the Gaussian likelihood with fixed variance is guaranteed to give the same posterior maximum as the true likelihood, for a flat prior \citep{Carron2013}. \cite{Hamimeche2008} investigated the extension to correlated fields and found that while the exactness of this relation does not hold, the Gaussian likelihood will still return the correct posterior maximum as long as the fiducial model is proportional to the model which maximises the likelihood. It is argued in that paper that for models which vary smoothly with $\ell$, this will often hold approximately even if it does not hold exactly.

The left panel of \autoref{Fig:postmax} shows the distribution of posterior maxima obtained from the two likelihoods for the 27\,000 realisations. The distributions are almost indistinguishable. The right panel shows the per-realisation difference between the posterior maximum returned by the two likelihoods. The Gaussian returns the correct maximum for 95.7 per cent of the realisations, and for the remainder it is wrong by no more than four grid points, which have a size of $\Delta w_0 = 1.25 \times 10^{-3}$. This demonstrates that -- as predicted in \cite{Hamimeche2008} -- the maximum-posterior property of the Gaussian likelihood holds to a very good approximation in practice for correlated fields.

\subsubsection{Posterior mean and standard deviation}

\begin{figure*}
\centering
\includegraphics[width=.95\textwidth]{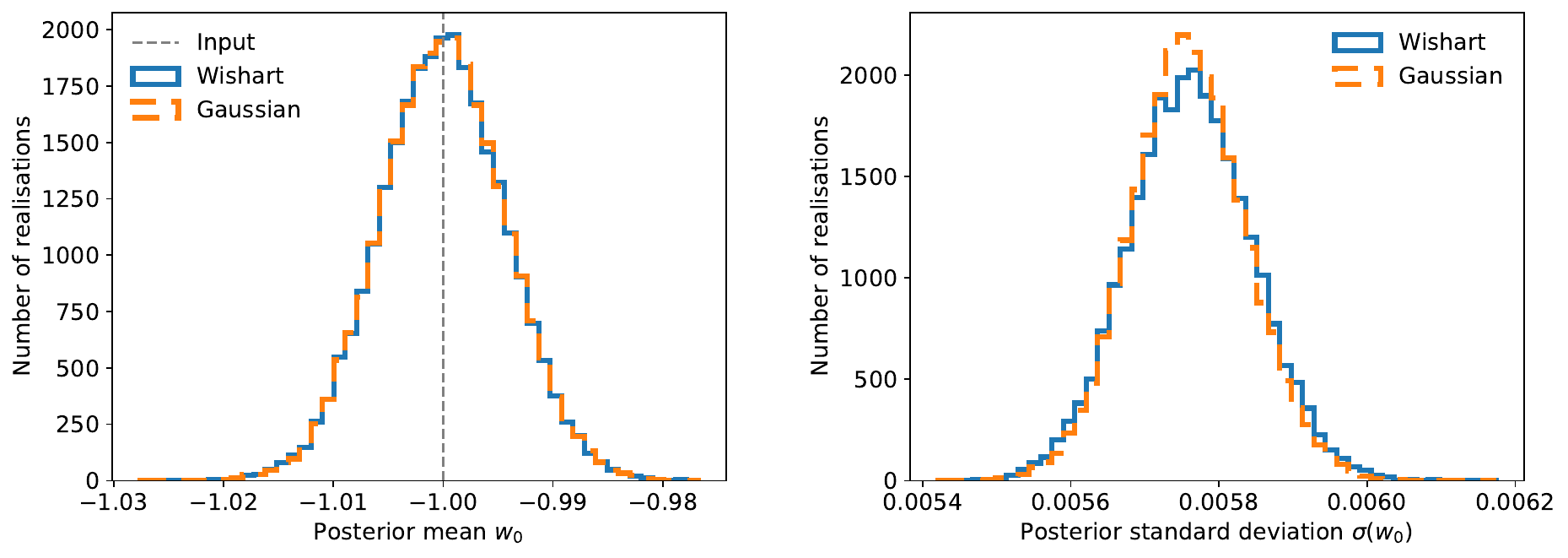}
\caption{\textit{Left:} Distribution of posterior means returned by the Gaussian likelihood compared to the true, Wishart likelihood. \textit{Right:} Distribution of posterior standard deviations for the two likelihoods.}
\label{Fig:postmeanstd}
\end{figure*}

Along with the posterior maximum, two other summary statistics 
for which it is perhaps most important for an approximate likelihood to return accurate values are the posterior mean and standard deviation. Unlike the posterior maximum, there is no general property of the Gaussian likelihood which says that it should return approximately correct values of these quantities. However, this appears to be the case on average: \autoref{Fig:postmeanstd} shows the distribution of posterior means (left panel) and standard deviations (right panel) for the Gaussian likelihood compared to the true, Wishart likelihood. The distributions of means are almost indistinguishable. The distributions of standard deviations are very similar, though there is some visible discrepancy. On further investigation we found that the Gaussian overestimates the standard deviation on realisations for which the true standard deviation is low (relative to its average over all realisations) and underestimates the standard deviation on realisations for which the true standard deviation is high. This effect has a magnitude of order 1 per cent of the true standard deviation. This is highly likely to be an acceptable level of inaccuracy, and is also expected to be smaller still when $\lmax$ is higher than the value of 100 used here. 

\subsection{Full sky: Posterior contours}
\label{Sec:fs_contours}

Cosmological parameter constraints are often visualised using two-dimensional contour plots, with the contours representing particular confidence intervals. Here we test the accuracy of the Gaussian likelihood in this regard. We use a single mock observation, produced following the method described in \autoref{Sec:fs_method_obs} with $\lmax = 2000$. This realisation was produced at random, but we have checked our results with different realisations and all give identical results in terms of level of agreement between the two likelihoods. As is the case in a real experiment, the posterior constraints are not centred on the input cosmology due to the sizeable contribution from cosmic variance inherent in a single realisation.

In most cases we perform a two-parameter likelihood analysis in ($w_0$, $w_a$) to keep computational costs down, but we also provide a three-parameter example to demonstrate that marginalisation over a third parameter does not affect the level of agreement between likelihoods. All two-dimensional posteriors are presented in terms of 1--3$\sigma$ contours, using the shorthand convention (deriving from the univariate Gaussian distribution) that 1, 2 and 3$\sigma$ represent 68.3, 95.4 and 99.7 per cent confidence. 

\subsubsection{Baseline setup}

\begin{figure*}
\centering
\includegraphics[width=.95\textwidth]{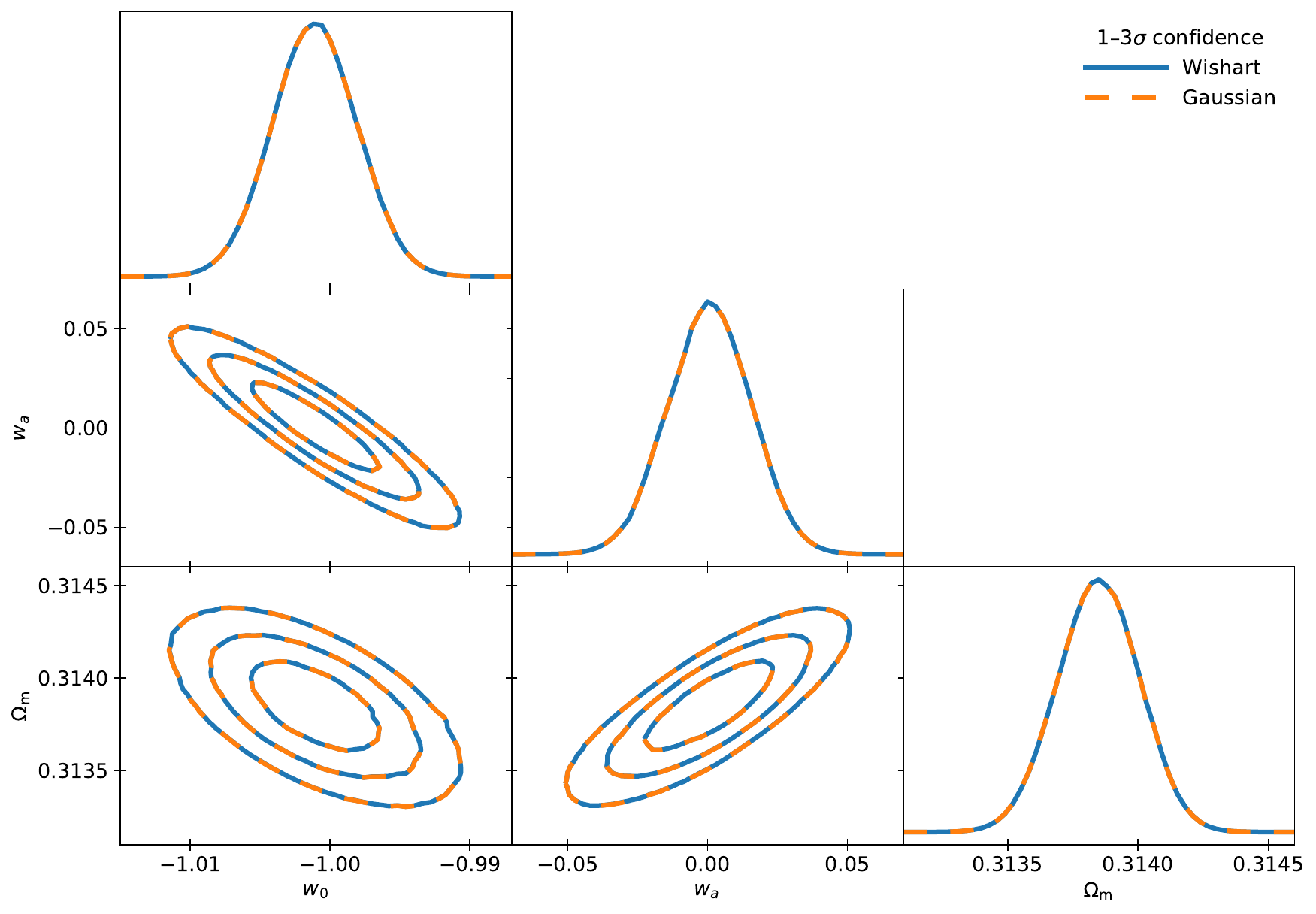}
\caption{Two- and one-dimensional marginalised posteriors from a three-parameter likelihood analysis using the Gaussian likelihood compared to the true, Wishart likelihood.}
\label{Fig:3dpost}
\end{figure*}

Our baseline full-sky test setup is as follows. In \autoref{Sec:robustness} we test the sensitivity of our results to the details of this setup. 
\begin{enumerate}
\item Five redshift bins (see \autoref{Sec:fs_method_theory}), each with galaxy number overdensity and shear $E$-mode fields;
\item All 55 3$\times$2pt power spectra between these ten fields; i.e., galaxy--galaxy, shear--shear and galaxy--shear;
\item Gaussian noise assuming a \textit{Euclid}-like number density evenly split between bins (see \autoref{Sec:fs_method_obs});
\item Multipole range $2 \leq \ell \leq 2000$;
\item Fiducial cosmology for Gaussian covariance equal to the true input cosmology used to generate the mock observation.
\end{enumerate}

\autoref{Fig:3dpost} shows two- and one-dimensional marginalised posterior distributions obtained from a three-parameter likelihood analysis with the Wishart and Gaussian likelihoods. The results from the two likelihoods are visually indistinguishable, showing that under our baseline setup the Gaussian likelihood is sufficiently accurate.

\subsubsection{Robustness to deviation from baseline setup}
\label{Sec:robustness}

It is important to check that the impressive degree of accordance between the Wishart and Gaussian likelihoods in \autoref{Fig:3dpost} is not a result of any specific choices made in the baseline setup outlined above. We now test the robustness of these results to deviations from this baseline setup. For these tests we perform a two-parameter likelihood analysis with other parameters fixed.

\paragraph*{Fiducial cosmology}

\begin{figure}
\includegraphics[width=\columnwidth]{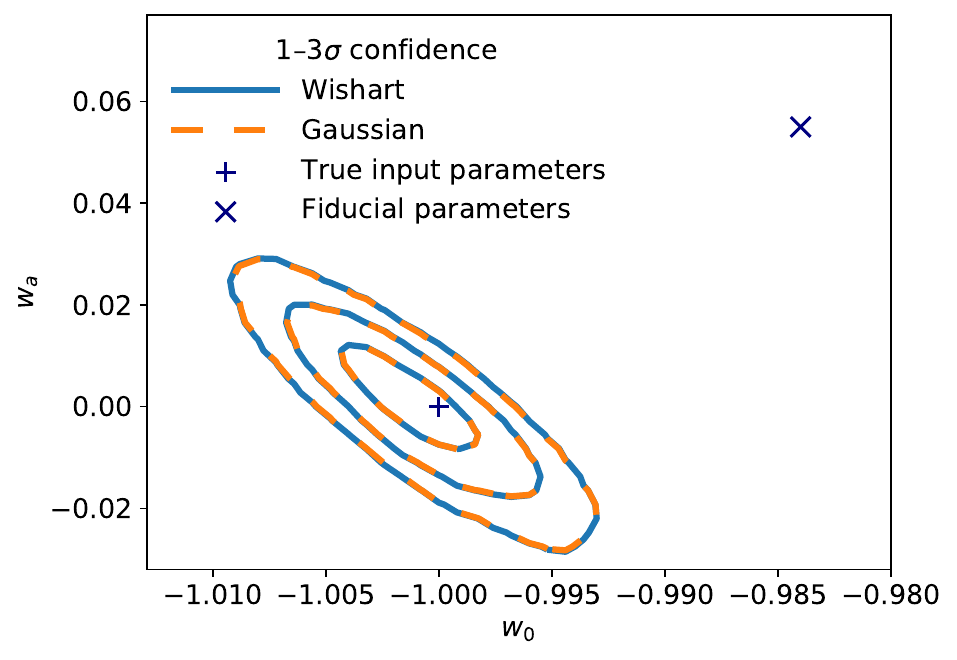}
\caption{Posterior distribution of $w_0$ and $w_a$ with other parameters fixed, where the fiducial cosmology used to evaluate the Gaussian covariance is excluded at high confidence.}
\label{Fig:fid_params}
\end{figure}

The Gaussian likelihood with fixed covariance requires choosing a fiducial cosmology at which to evaluate the covariance. In the baseline setup, we chose the fiducial cosmology to be equal to the true input cosmology that was used to generate the mock observation. In a real analysis this would not be possible, since the true cosmology is unknown. To model this effect, we have repeated the analysis with the fiducial cosmology chosen to be distant from the true cosmology. \autoref{Fig:fid_params} shows one example, for which the fiducial cosmology is excluded at more than $10 \sigma$ and yet this does not appear to affect the accuracy of the Gaussian likelihood. We have found that the accuracy does eventually diminish, but only when the fiducial cosmology and the true cosmology are unrealistically far apart (e.g. using a fiducial $w_0 = -0.2$ and a true $w_0 = -1.0$). Even in this case, it is the size and shape of the posterior distribution that is affected, much more than its location. In any real analysis, if the fiducial cosmology were excluded at high confidence then the analysis should be repeated with a fiducial cosmology consistent with the data. Therefore, even if posterior parameter constraints in the initial case were inaccurate due to the choice of fiducial cosmology, they would converge onto the correct constraints through this process.

\paragraph*{$\bm{\ell}$ range}

\begin{figure}
\includegraphics[width=\columnwidth]{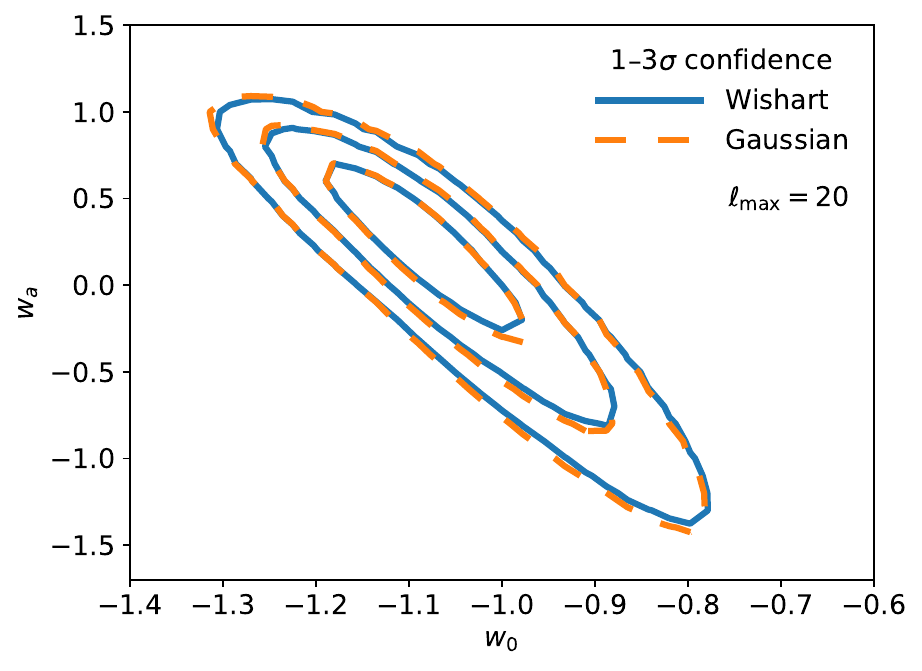}
\caption{Posterior distribution of $w_0$ and $w_a$ with other parameters fixed, with only $\ell = $ 2--20 included in the likelihood.}
\label{Fig:lmax20}
\end{figure}

The Gaussian likelihood should be expected to perform best at high $\ell$, as the true likelihood gradually tends to Gaussian by the Central Limit Theorem as more $\alm$s contribute to each $\Cl$ estimate (see \autoref{Sec:cut_sky}). Therefore, as $\lmax$ is reduced for a constant $\lmin$, the accuracy of the Gaussian likelihood should decrease. We observe this expected behaviour, but it is surprisingly weak. \autoref{Fig:lmax20} shows the posterior distribution obtained with $\lmax = 20$. While there is some disagreement between the two sets of contours, the Gaussian likelihood is still very clearly able to recover the non-Gaussian shape of the true posterior.

\paragraph*{Noise}

\begin{figure}
\includegraphics[width=\columnwidth]{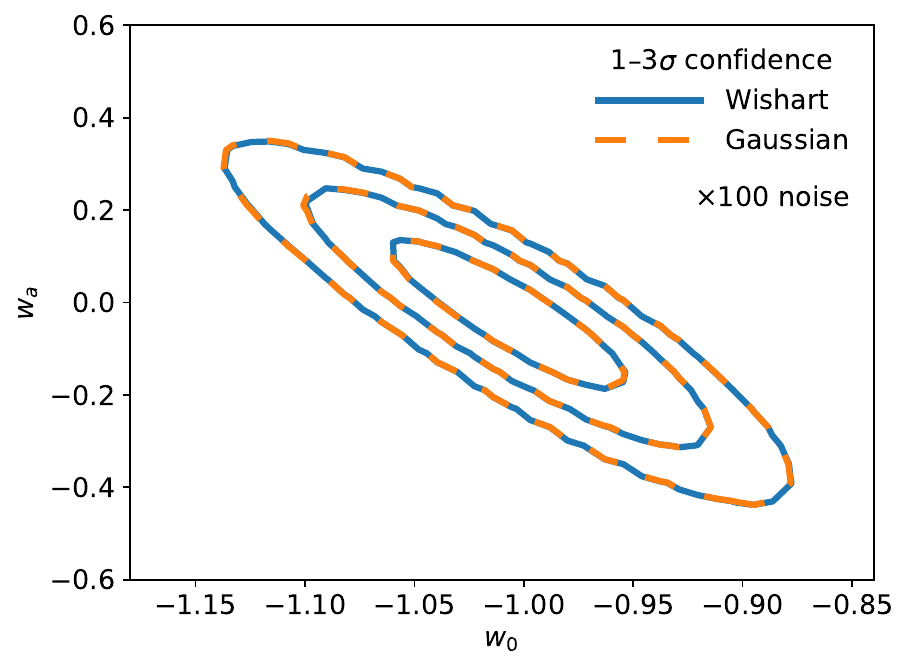}
\caption{Posterior distribution of $w_0$ and $w_a$ with other parameters fixed, with noise at $100 \times$ \textit{Euclid}-like levels.}
\label{Fig:x100noise}
\end{figure}

Even under the assumption of Gaussian fields and Gaussian noise, the level of noise has a theoretical impact on the accuracy of the Gaussian likelihood. This is because the noise power spectrum is flat, while the signal power spectra all decrease with $\ell$. For any noise level, there is a threshold $\ell$ above which the noise dominates the signal. Increasing the noise level decreases this threshold, meaning that a greater fraction of the overall constraining power of the data comes from lower $\ell$. Therefore, increasing the noise level relatively upweights the contribution of lower $\ell$, which -- as discussed above -- is the subset of the data for which the Gaussian likelihood should perform worst. However, this does not appear to have a noticeable effect for realistic noise levels. \autoref{Fig:x100noise} shows the posterior distributions obtained with $100 \times$ \textit{Euclid}-like noise, achieved by assuming a total galaxy number density of $0.3 / \text{arcmin}^2$. 
We have also tried decreasing noise (and switching it off entirely for a reduced 2-redshift-bin setup), though for Gaussian fields and Gaussian noise this should not decrease the accuracy of the Gaussian likelihood; rather, it should increase following the inverse of the above argument. In both cases this did not lead to any visible discrepancy between the two posteriors.

We also tried varying other aspects of the baseline setup, including testing with a single power spectrum and testing other parameter combinations ($\Omega_\text{m}$--$\sigma_8$, $w_0$--$n_\text{s}$, $\Omega_\text{m}$--$\sigma_8$--$w_0$) but none of these made any visible difference to the level of accordance between the two likelihoods. One change that did make a significant difference was allowing the covariance matrix in the Gaussian likelihood to vary across parameter space rather than being fixed, confirming that this aspect is crucial to the accuracy of the Gaussian likelihood. With the covariance fixed, we conclude that the Gaussian likelihood is sufficiently accurate for full-sky power spectra from Gaussian fields.

\section{Cut-sky likelihood}
\label{Sec:cut_sky}

We cannot necessarily assume that the accuracy of the Gaussian likelihood on the full sky will extend to the cut sky, where the situation is more complicated. Although the exact cut-sky likelihood under the assumption of Gaussian fields is known \citep{Upham2019a}, it is only feasible to use in its exact form in specific low-dimensional cases. 
On the full sky, all $\alm$s are independent, and are identically distributed for a given $\ell$. The introduction of a mask mixes the $\alm$s \citep{Lewis2001, Brown2005}, breaking these two properties. For an exact treatment, it becomes necessary to keep track of the relationship between all $\alm$s, which quickly becomes impossible for a high-dimensional analysis such as this one.

This motivates an alternative approach, which we take here: after having tested the accuracy of the Gaussian likelihood on the full sky, where the exact likelihood is both known and tractable, we now carefully consider the ways in which a sky cut might decrease this accuracy. To do this, we compare the non-Gaussianity of the cut-sky likelihood to the non-Gaussianity of the full-sky likelihood. We utilise Sklar's theorem, which states that any multivariate probability distribution may be separated into its marginal distributions and its dependence structure \citep{Sklar1959}. The dependence structure is called the copula, though for clarity we choose not to use this term here to avoid confusion with the method of forming an approximate joint distribution by combining separate approximations for marginals and the copula, commonly using a Gaussian copula \citep[see][for discussion of this method in a cosmological context]{Benabed2009, Sato2010, Sato2011}. We therefore consider the non-Gaussianity of the marginal distributions in \autoref{Sec:ma_marginals} and the non-Gaussianity of the dependence structure in \autoref{Sec:ma_dependence}, in each case comparing between the full-sky and cut-sky likelihoods.

\subsection{Cut sky: Effect on marginal distributions}
\label{Sec:ma_marginals}

\begin{figure}
\includegraphics[width=\columnwidth]{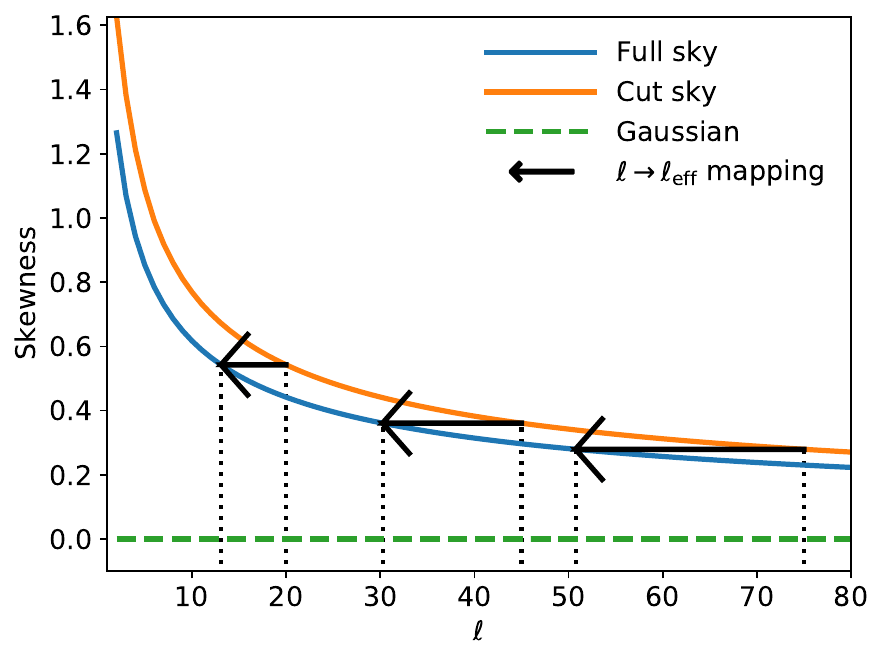}
\caption{Skewness of the full-sky and cut-sky marginal auto-$C_\ell$ distribution as a function of $\ell$, where the cut-sky result is for a \textit{Euclid}-like mask. The arrows demonstrate the $\ell \rightarrow \leff$ mapping described in \autoref{Sec:ma_marginals_impact}.}
\label{Fig:skewness}
\end{figure}

We focus our investigation into the non-Gaussianity of marginal distributions on auto-spectra, which by their positive-definite nature are much more non-Gaussian than cross-spectra on both the full and cut sky \citep{Percival2006, Upham2019a}. We quantify non-Gaussianity using the skewness and excess kurtosis, following \cite{Lin2020} and \cite{DiazRivero2020}, since both vanish for a Gaussian distribution. These can be written in terms of the mean $E \left[ X \right]$ and standard deviation $\text{Std} \left( X \right)$ of a random variable $X$ as
\begin{align}
&\text{Skew} \left( X \right) = 
E \left[ \left( \frac{X - E \left[ X \right]}
{ \text{Std} \left( X \right) } \right)^3 \right]; \\
&\text{Ex kurt} \left( X \right) = 
E \left[ \left( \frac{X - E \left[ X \right]}
{ \text{Std} \left( X \right) } \right)^4 \right] - 3.
\end{align}
The marginal distribution of a full-sky auto-$C_\ell$ (the diagonal elements of \autoref{Eqn:obs_cl_matrix}) is a gamma distribution, which under the ($k$, $\theta$) parametrisation has PDF
\begin{equation}
f_\Gamma \left( x | k, \theta \right) = 
\frac{x^{k - 1} \exp \left[ - x / \theta \right]}
{\Gamma (k) \theta^k},
\end{equation}
where $\Gamma$ is the gamma function. This distribution has skewness and excess kurtosis
\begin{equation}
\text{Skew} \left( X \right) = \frac{2}{\sqrt{k}};
\quad\quad
\text{Ex kurt} \left( X \right) = \frac{6}{k}.
\end{equation}
The full-sky likelihood corresponds to parameter values \citep{Percival2006, Hamimeche2008, Sellentin2018a}
\begin{equation}
\widehat{C}_\ell \sim \Gamma \left( 
k = \frac{2 \ell + 1}{2},~ 
\theta = \frac{2 C_\ell}{2 \ell + 1} \right).
\end{equation}
Both the skewness and kurtosis therefore depend only on $\ell$, and are both power laws in $2 \ell + 1$:
\begin{align}
&\text{Skew} \left( \widehat{C}_\ell \right) 
= \sqrt{8} \left[ 2 \ell + 1 \right]^{- 1/2};
\label{Eqn:skew_fs} \\
&\text{Ex kurt} \left( \widehat{C}_\ell \right) 
= 12 \left[ 2 \ell + 1 \right]^{-1}.
\label{Eqn:kurt_fs}
\end{align}
The skewness and excess kurtosis of the cut-sky likelihood may be derived from the pseudo-$C_\ell$ marginal characteristic function (CF),
\begin{equation}
\varphi_{\widetilde{C}_\ell} \left( t \right) = \prod_j 
\left( 1 - 2i \lambda_j t \right)^{-1/2},
\label{Eqn:marg_cf}
\end{equation}
where $\{ \lambda_j \}$ are the eigenvalues of
$\mathbfss{M} \bm{\Sigma}$, the product of the pseudo-$a_{\ell m}$ covariance matrix $\bm{\Sigma}$ with $\mathbfss{M}$, the selection matrix picking out the relevant elements of $\bm{\Sigma}$ for the $C_\ell$ in question \citep[see][]{Upham2019a}. \autoref{Eqn:marg_cf} may be identified as a product of gamma distribution CFs:
\begin{equation}
\varphi_\Gamma \left( t \right) = 
\left( 1 - \theta i t \right)^{-k},
\end{equation}
each with parameters $k = 1/2$, $\theta = 2 \lambda_j$. Since the CF of a sum of independent random variables is equal to the product of the individual CFs, it follows that the marginal distribution of a pseudo-$C_\ell$ estimate is identical to that of a sum of independent gamma-distributed variables. This allows the calculation of the cut-sky skewness and excess kurtosis in terms of the eigenvalues $\lambda_j$ of $\mathbfss{M} \bm{\Sigma}$:
\begin{equation}
\text{Skew} \left( \widetilde{C}_\ell \right)
= \frac{ 2^{3/2} \sum_j \lambda_j ^3 }
{ \left[ \sum_j \lambda_j^2 \right]^{3 / 2}};
\quad\quad
\text{Ex kurt} \left( \widetilde{C}_\ell \right)
= \frac{12 \sum_j \lambda_j^4}
{\left[ \sum_j \lambda_j^2 \right]^2}.
\label{Eqn:skewkurt_ma}
\end{equation}
\autoref{Fig:skewness} shows the full- and cut-sky skewness as a function of $\ell$, up to $\ell = 80$, for a \textit{Euclid}-like mask incorporating the survey footprint and a bright star mask ($f_\text{sky} = 30.7$ per cent). Both curves are smoothly decreasing, with the cut-sky skewness systematically higher. The kurtosis exhibits a similar behaviour. The marginal distributions of the likelihood, therefore, are more non-Gaussian on the cut sky than on the full sky. We now investigate the impact of this additional non-Gaussianity.

\subsubsection{Impact of additional non-Gaussianity}
\label{Sec:ma_marginals_impact}

\begin{figure*}
\centering
\includegraphics[width=.95\textwidth]{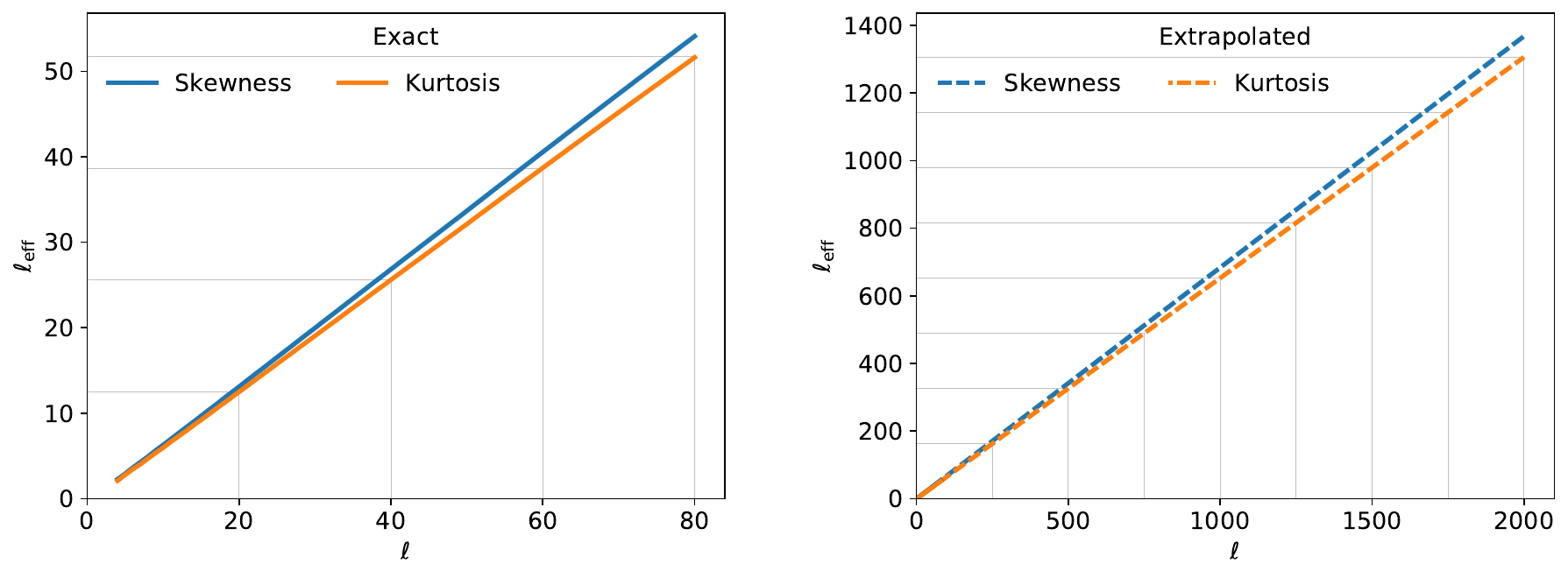}
\caption{\textit{Left:} $\ell \rightarrow \leff$ mapping derived from equating the skewness (blue) and excess kurtosis (orange) of the full-sky and cut-sky likelihoods. \textit{Right:}~Extrapolated to $\ell = 2000$.}
\label{Fig:leff_map}
\end{figure*}

\begin{figure*}
\centering
\includegraphics[width=.95\textwidth]{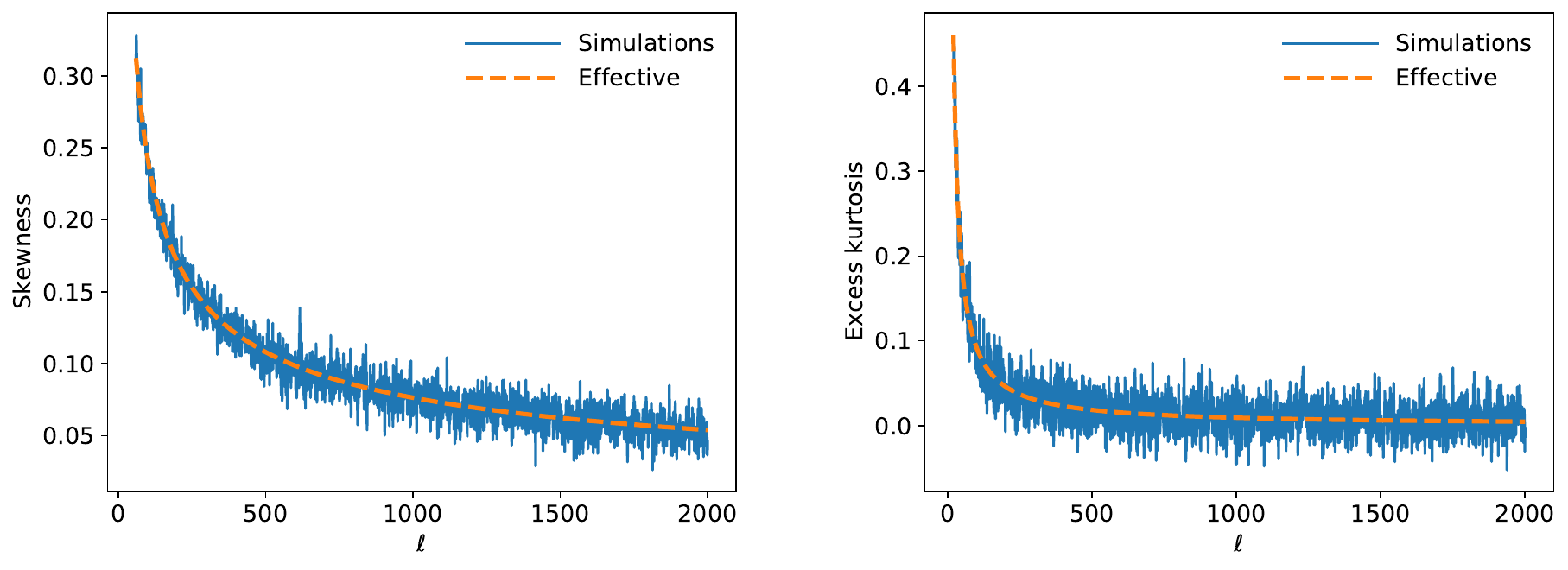}
\caption{Validation of the extrapolation in \autoref{Fig:leff_map} against 63\,100 simulated cut-sky realisations.}
\label{Fig:leff_validation}
\end{figure*}

\begin{figure}
\includegraphics[width=\columnwidth]{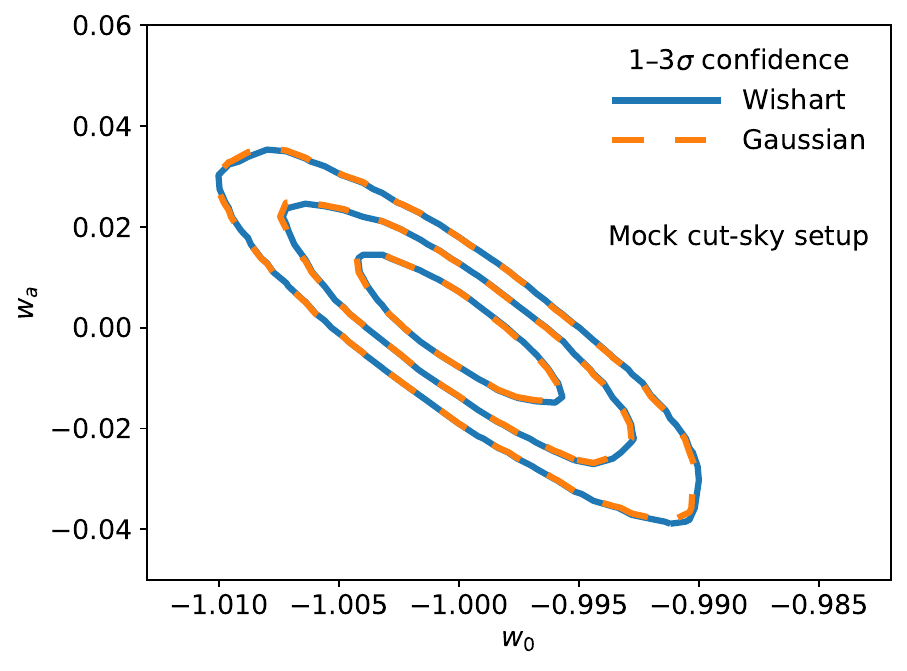}
\caption{Posterior distributions obtained from a mock observation designed to have the same amount of non-Gaussianity in its marginal distributions as the cut-sky likelihood.}
\label{Fig:leff_post}
\end{figure}

We take advantage of the fact that both skewness and kurtosis are higher on the cut sky and that both decrease smoothly with $\ell$ to define an ``effective $\ell$'', $\leff$, for each $\ell$ by equating the full- and cut-sky skewness, and the same for kurtosis. This process is demonstrated by the arrows in \autoref{Fig:skewness}. This $\ell \rightarrow \leff$ mapping is shown in the left panel of \autoref{Fig:leff_map}, and turns out to be perfectly linear for both skewness and kurtosis. This is unexpected, since it is not apparent from the expressions for the full- and cut-sky skewness and excess kurtosis in Equations \ref{Eqn:skew_fs}--\ref{Eqn:kurt_fs} and \ref{Eqn:skewkurt_ma}. However, we have found that it holds for all ten auto-spectra in our setup. It appears that the cut-sky skewness and excess kurtosis as a function of $\ell$ are simply linear transformations of their full-sky counterparts, with the slope of the transformation depending on the details of the mask.

In the right panel of \autoref{Fig:leff_map} we extrapolate this linear mapping to $\ell = 2000$. Since this is a large extrapolation, we verify it in \autoref{Fig:leff_validation} by comparing to the sample skewness and kurtosis from 63\,100 simulated cut-sky realisations of a single field. It is clearly an excellent fit.

Finally, we test the impact of this additional non-Gaussianity of the marginal distributions on the cut sky by applying an adjusted Wishart likelihood, which has the correct amount of cut-sky non-Gaussianity in its marginal distributions, by replacing $\ell$ in the likelihood with $\leff$. We use the kurtosis mapping, since it gives a lower $\leff$ for a given $\ell$ (\autoref{Fig:leff_map}) and is therefore a more pessimistic choice. The adjusted likelihood replaces \autoref{Eqn:wishart} with
\begin{equation}
\widehat{\mathbfss{C}}'_\ell \sim 
\mathcal{W} \left( \nu = 2 \leff + 1, 
\mathbfss{V} = \frac{\mathbfss{C}_\ell}{2 \leff + 1} \right).
\label{Eqn:adj_wishart}
\end{equation}
Note that each observed $\ell$ still depends on the same $\ell$ in the theory power spectra. This means that each $\Cl$ will retain the correct sensitivity to cosmological parameters, enabling us to test the impact of an increased amount of non-Gaussianity for the same cosmological constraining power. The marginal distributions of the auto-$\Cl$s in $\widehat{\mathbfss{C}}'_\ell$ are gamma distributions, with parameters
\begin{equation}
\widehat{C}'_\ell \sim \Gamma \left( 
k = \frac{2 \leff + 1}{2},~ 
\theta = \frac{2 C_\ell}{2 \leff + 1} \right),
\end{equation}
and therefore -- from Equations \ref{Eqn:skew_fs}--\ref{Eqn:kurt_fs} -- will have the same amount of skewness and excess kurtosis as the cut-sky likelihood (in fact a slightly higher amount of skewness, since we use the kurtosis $\ell \rightarrow \leff$ mapping). 

The corresponding Gaussian likelihood has the same mean and covariance as the adjusted Wishart likelihood. Its mean, therefore, is unchanged from the full-sky case, while the covariance is
\begin{equation}
\text{Cov} \left(
\widehat{C}_\ell^{\alpha \beta},
\widehat{C}_{\ell'}^{\gamma \varepsilon}
\right)
= \frac{\delta_{\ell \ell'}}{2 \leff + 1} \left(
C_\ell^{\alpha \gamma} C_\ell^{\beta \varepsilon}
+ C_\ell^{\alpha \varepsilon} C_\ell^{\beta \gamma} \right).
\end{equation}

We generated a mock observation following \autoref{Eqn:adj_wishart} by sampling directly from the Wishart distribution using the SciPy implementation of the Wishart variate generating algorithm from \cite{Smith1972}. Using this observation, we conduct a two-parameter likelihood analysis with the adjusted Wishart and Gaussian likelihoods. The resulting posterior distribution is shown in \autoref{Fig:leff_post}. There is very good agreement between the two likelihoods. Although small deviations are visible, this is highly likely to be an acceptable level of inaccuracy. We conclude that the additional non-Gaussianity in the marginal distributions of the cut-sky likelihood compared to the full-sky likelihood is insufficient to introduce significant inaccuracy into the results obtained using a Gaussian likelihood.

\subsection{Cut sky: Effect on dependence structure}
\label{Sec:ma_dependence}

\begin{figure}
\includegraphics[width=\columnwidth]{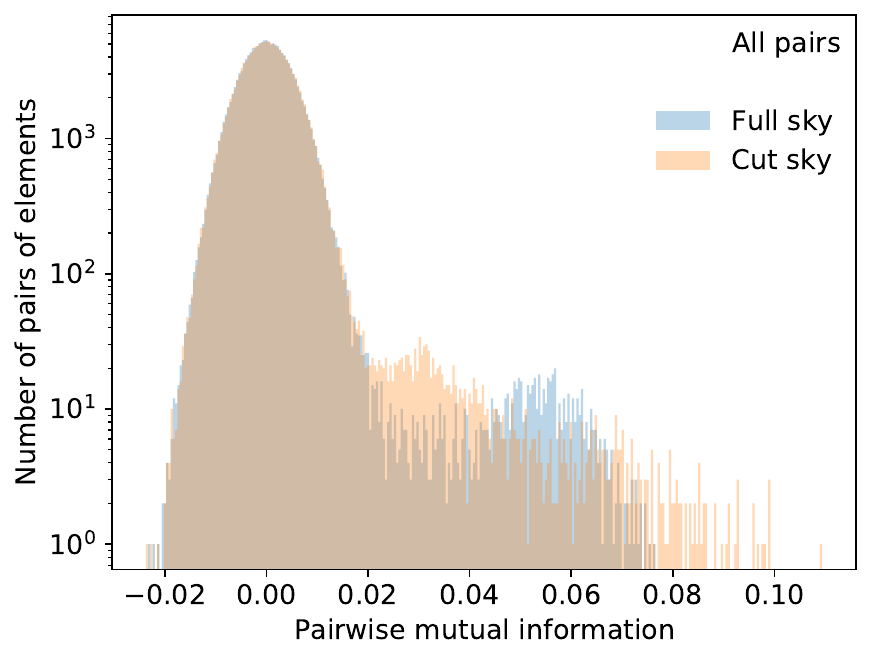}
\caption{Comparison of non-Gaussian dependence between full-sky and cut-sky likelihoods. Non-Gaussian dependence is quantified by pairwise mutual information after whitening. The mass centred around zero on the $x$-axis represents Gaussian dependence, with the tail (here $\gtrsim 0.02$) representing non-Gaussian dependence.}
\label{Fig:mi_all}
\end{figure}

\begin{figure*}
\centering
\includegraphics[width=.95\textwidth]{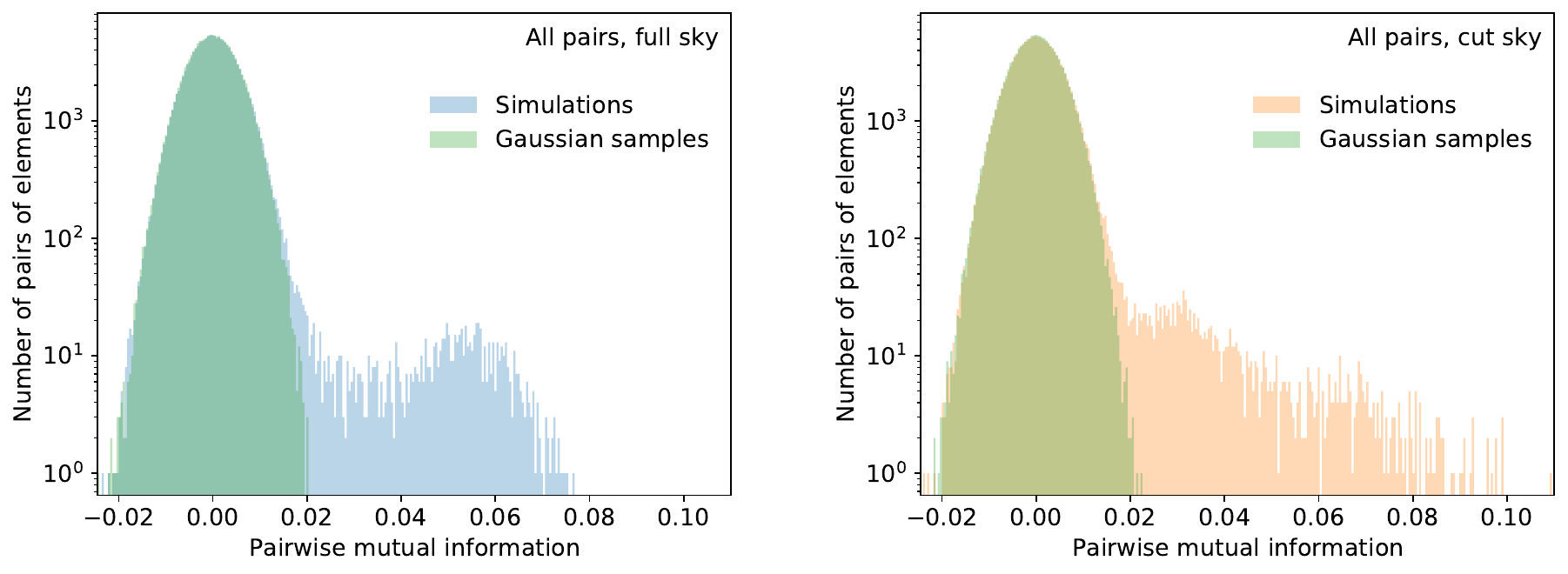}
\caption{Comparison of non-Gaussian dependence between full-sky (left) and cut-sky (right) likelihoods. Non-Gaussian dependence is quantified by pairwise mutual information after whitening. In both panels we show the corresponding distribution for pure Gaussian dependence, which is centred around zero on the $x$-axis. The tail (here $\gtrsim 0.02$) represents non-Gaussian dependence.}
\label{Fig:mi_vs_gauss}
\end{figure*}

\begin{figure*}
\centering
\includegraphics[width=.95\textwidth]{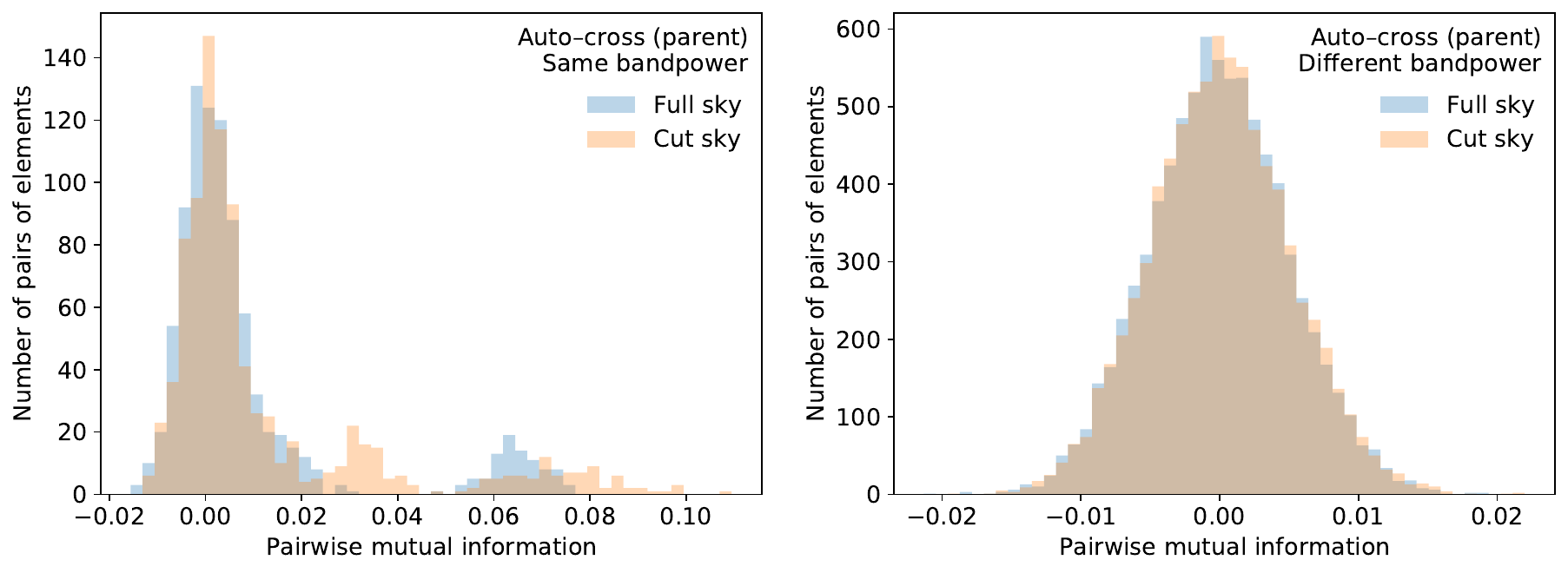}
\caption{Comparison of non-Gaussian dependence between full-sky and cut-sky likelihoods, for two specific populations of data element pairs. The left panel shows pairs containing the equivalent bandpower of one cross-spectrum and one of its `parent' auto-spectra (e.g. $P_b^{12}$ and $P_b^{11}$). The right panel is for different-bandpower pairs in the same combinations of spectra (e.g. $P_b^{12}$ and $P_{b'}^{11}$). Non-Gaussian dependence is quantified by pairwise mutual information after whitening. The bulk centred around zero on the $x$-axis represents Gaussian dependence, with the tail representing non-Gaussian dependence. The absence of this tail in the right panel indicates that those pairs exhibit only Gaussian dependence.}
\label{Fig:mi_parent}
\end{figure*}

\begin{figure*}
\centering
\includegraphics[width=.95\textwidth]{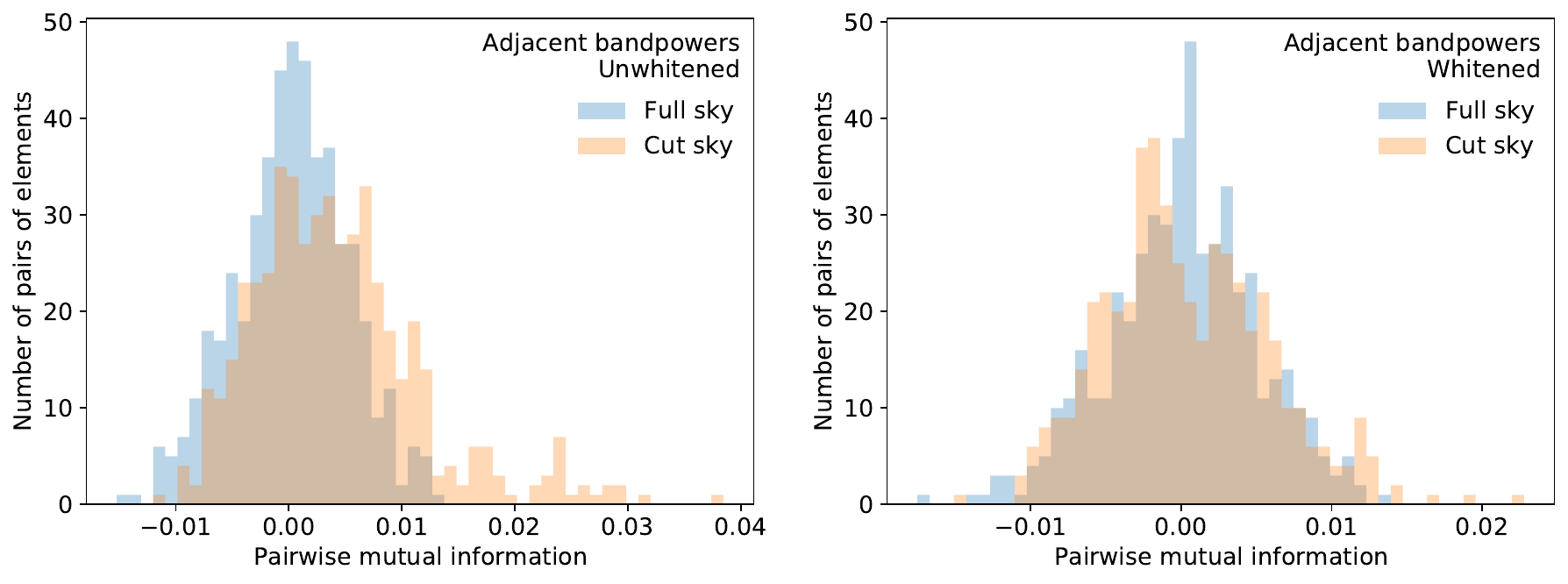}
\caption{Full-sky and cut-sky distributions of pairwise mutual information (MI) before (left) and after (right) pairwise whitening, for pairs of adjacent bandpowers in the same spectrum (e.g. $P_b^{12}$ and $P_{b + 1}^{12}$). Prior to whitening, MI captures all dependence; in this case there is a substantial excess of dependence in the cut-sky likelihood. Whitening removes Gaussian dependence such that after whitening, MI only captures non-Gaussian dependence; in this case there is little or no excess dependence in the cut-sky likelihood.  This demonstrates that the additional dependence between adjacent bandpowers induced by a mask is mostly or wholly Gaussian.}
\label{Fig:mi_adj_bp}
\end{figure*}

To study the cut-sky dependence structure we rely on simulations. We generated 50\,000 simulated observations following the method described in \autoref{Sec:fs_method_obs} with two differences: first we measured observed power spectra for each realisation both before and after multiplication at the map level by the \textit{Euclid}-like mask; we then formed 10 logarithmically spaced bandpowers from $\ell =$ 2 to 2000, weighted following Equation 20 of \cite{Hivon2002}.

\subsubsection{Mutual information}

We quantify dependence between two data elements using mutual information (MI). MI is defined as the Kullback--Leibler (KL) divergence $D_\text{KL}$ of the joint distribution of two variables from the product of their marginal distributions,
\begin{equation}
I \left( X, Y \right)
= D_\text{KL} \left( P_{\left( X, Y \right)} ~\middle| \middle|~ 
P_X \otimes P_Y \right),
\end{equation}
where the KL divergence for continuous distributions is 
\begin{equation}
D_\text{KL} \left( P_{\left( X, Y \right)} ~\middle| \middle|~ 
P_X \otimes P_Y \right)
= \iint dx\,dy ~ p \left( x, y \right)
\log \left( \frac{ p \left(x, y \right) }
{p \left(x \right) p \left( y \right)} \right).
\label{Eqn:kl_div}
\end{equation}
If $X$ and $Y$ are independent, their joint distribution factorises and the MI vanishes. If they are not independent, they will have a positive MI. In practice, however, MI estimation from a finite number of samples may return a negative value.

To isolate non-Gaussian dependence, we first apply a whitening procedure to remove linear correlations. Linear correlations are those which are fully described by a covariance (or equivalently, correlation) matrix. Since the dependence structure in a multivariate Gaussian distribution is also fully described by its covariance matrix -- such that the components of a multivariate Gaussian with diagonal covariance are independent -- removing linear correlations removes all Gaussian dependence. 
This whitening follows the same process as \cite{Sellentin2018}, \cite{Sellentin2018a}, \cite{DiazRivero2020} and \cite{Louca2020}: each pair of data elements is whitened separately using a Cholesky whitening procedure followed by a mean subtraction. The result is a whitened pair having a mean of zero and a covariance matrix of the identity matrix. We whiten each pair separately so that pairs are still identifiable, allowing us to study the behaviour of pairs with specific relationships.

For each whitened pair, we estimate MI using the Non-parametric Entropy Estimation Toolbox (NPEET)\footnote{\href{https://github.com/gregversteeg/NPEET}{https://github.com/gregversteeg/NPEET}} \citep{VerSteeg2014}. The NPEET MI estimator implements a $k$-nearest neighbours method described in \cite{Kraskov2004}. We use the default of $k = 3$ and log base 2 in \autoref{Eqn:kl_div}.

\autoref{Fig:mi_all} shows the distribution of pairwise whitened MI compared between full-sky and cut-sky bandpowers. Most of the pairs of elements are found in the part of the distribution centred around zero, indicating no detected non-Gaussian dependence. This is more clearly seen in \autoref{Fig:mi_vs_gauss}, in which each of the full-sky and cut-sky samples is compared to an equivalent sample drawn from a multivariate Gaussian distribution having the same mean and covariance. Non-Gaussian dependence is exhibited by the pairs of elements found in the tail, in this case with MI $\gtrsim 0.02$. This tail contains only a small fraction of pairs in both cases, but with a slight excess for the cut-sky sample: 0.93 per cent of cut-sky pairs have MI $> 0.02$, compared to 0.65 per cent of full-sky pairs. This is also evident in the small visible excess of cut-sky pairs in \autoref{Fig:mi_all}.

To investigate the origin of this small excess in non-Gaussian dependence for the cut-sky sample relative to the full-sky sample, we split each sample into different pair populations, corresponding to particular relationships between data elements. We find that non-Gaussian dependence is almost exclusively found in pairs containing the same bandpower across correlated fields. The strongest such case is shown in the left panel of \autoref{Fig:mi_parent}, which shows pairs containing one bandpower from a cross-spectrum and the same bandpower from one of its `parent' auto-spectra, i.e. the auto-spectrum of one of the two fields between which the cross-spectrum is describing the correlation. While most pairs still appear consistent with zero, there is a significant tail of non-Gaussian dependence, which is slightly larger for the cut-sky sample. This tail is not found when looking at pairs of different bandpowers between the same two spectra, shown in the right panel of \autoref{Fig:mi_parent}. We find a similar behaviour in other same-bandpower pairs, which is strongest when the two spectra in the pair relate directly to the same underlying field; for example, two `sibling' cross-spectra which share one parent auto-spectrum. In most such pair populations, there is a slight excess of non-Gaussian dependence for the cut-sky sample.

In contrast, we find that the dependence between bandpowers in the same spectrum known to be induced by a cut sky in fact comprises almost purely linear correlations. This is shown in \autoref{Fig:mi_adj_bp}, which compares these pairs before and after the whitening process. The left panel shows the unwhitened result, which includes linear correlations, showing an expected cut-sky excess. After whitening, shown in the right panel, this excess is almost entirely removed.

\subsubsection{Impact of additional non-Gaussian dependence}

\begin{figure}
\includegraphics[width=\columnwidth]{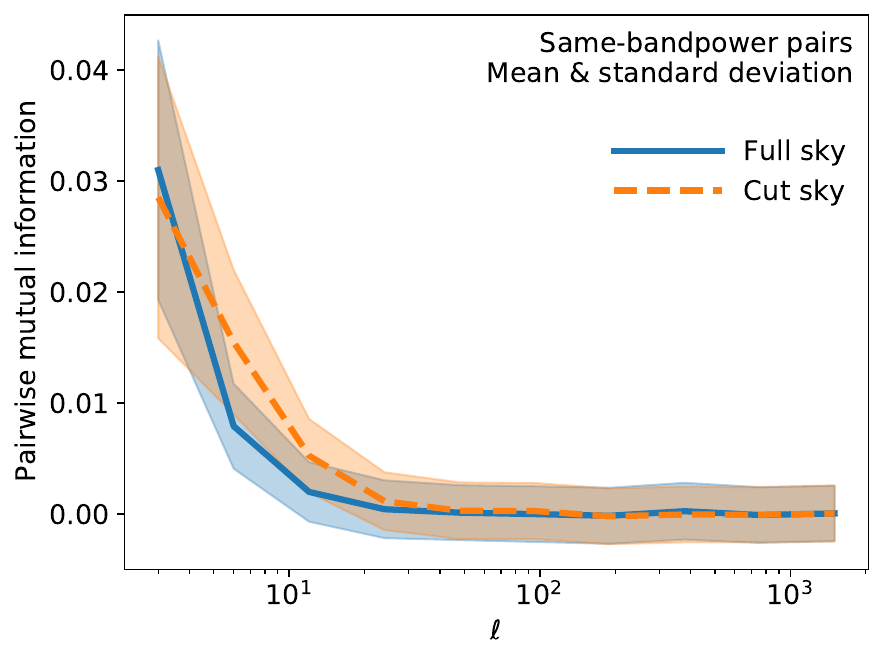}
\caption{Non-Gaussian dependence as a function of $\ell$, compared between full-sky and cut-sky likelihoods. Non-Gaussian dependence is quantified by pairwise mutual information after whitening, and is here averaged over all same-bandpower pairs in a given $\ell$ bin, with the shaded region containing one standard deviation.}
\label{Fig:mi_vs_l}
\end{figure}

In the above section we have shown that there is a small excess in non-Gaussian dependence in the cut-sky likelihood compared to the full-sky likelihood. As we did for the marginal distributions in \autoref{Sec:ma_marginals_impact}, we now investigate the potential impact of this additional non-Gaussianity on the accuracy of constraints obtained using the Gaussian likelihood.

On closer inspection we find that the increased non-Gaussian dependence in the cut-sky likelihood is in fact an $\ell$-dependent effect. This is demonstrated in \autoref{Fig:mi_vs_l}, which shows whitened MI as a function of $\ell$ for same-bandpower pairs, which as discussed above are those which exhibit non-Gaussian dependence. Non-zero MI appears to be restricted only to the lowest bandpowers, with a small excess for the cut sky. This $\ell$ dependence resembles that of the skewness and kurtosis of the marginal distributions, and implies that the mock cut-sky data vector and likelihood that we developed in \autoref{Sec:ma_marginals_impact} using the $\ell \rightarrow \leff$ mapping process should have higher MI -- indicating more non-Gaussian dependence -- than the full-sky data and likelihood tested in \autoref{Sec:fullsky}. We are able to conservatively estimate the average MI in the mock cut-sky setup by taking the full-sky MI sample and replacing the MI value of each same-bandpower pair at any $\ell$ with that of its corresponding $\leff$, interpolating the full-sky MI-vs.-$\ell$ curve shown in \autoref{Fig:mi_vs_l}. We leave the MI of different-bandpower pairs unchanged. This gives an average MI of $9.0 \times 10^{-4}$, which compares to $4.1 \times 10^{-4}$ for the full-sky sample and $5.2 \times 10^{-4}$ for the cut-sky sample. So the mock cut-sky sample and likelihood has roughly 80 per cent more non-Gaussian dependence than the true cut-sky sample and likelihood, and yet the resulting posterior distribution from the Gaussian likelihood in \autoref{Fig:leff_post} is still extremely accurate. Therefore, we conclude that the impact of additional non-Gaussian dependence on the cut sky is negligible.

\subsubsection{Transcovariance}
\label{Sec:transcov}

\begin{figure}
\includegraphics[width=\columnwidth]{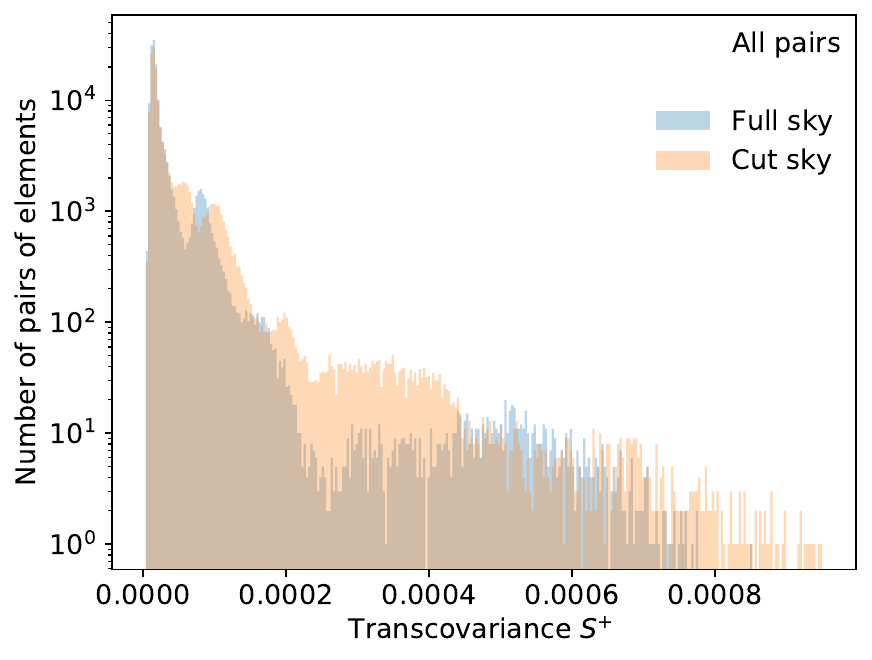}
\caption{Full-sky and cut-sky distributions of pairwise additive transcovariance after pairwise whitening. Transcovariance is an alternative measure of non-Gaussianity, discussed in \autoref{Sec:transcov}.}
\label{Fig:transcov}
\end{figure}

Transcovariance is a measure of non-Gaussianity of a distribution introduced in \cite{Sellentin2018} and subsequently used in \cite{Sellentin2018a}, \cite{Louca2020} and \cite{DiazRivero2020}. We will follow the latter three papers in considering only the additive transcovariance $S^+$, which is defined as
\begin{equation}
S^+ = \frac{1}{B} \sum_{b = 1}^B
\left[ \mathcal{H}_b - \mathcal{N}_b \left(0, 2 \right) \right]^2,
\end{equation}
where the sum is over the bins $b$ of a histogram $\mathcal{H}$ of the sum of two data elements after whitening, and $\mathcal{N} \left(0, 2 \right)$ is the expected histogram of a univariate Gaussian distribution with mean $0$ and variance $2$. 

$S^+$ is a measure of non-Gaussianity, because if the two data elements were bivariate Gaussian distributed, then their sum after whitening would be univariate Gaussian with mean $0$ and variance $2$, and so the expectation $\text{E} \left[ \mathcal{H}_b - \mathcal{N}_b \left(0, 2 \right) \right]$ would vanish. $S^+$ has sometimes been described as a measure of "non-Gaussian correlations", but in practice it is sensitive to non-Gaussianity of both the marginals and the dependence.
For example, if two data elements each had a non-Gaussian marginal distribution but their dependence was purely linear correlation (or more simply, if they were independent), then their dependence would vanish with the whitening procedure and yet they would return a non-zero $S^+$ value because their sum would not, in general, follow a Gaussian distribution. As such, $S^+$ is a holistic measure of non-Gaussianity, which is a useful property in many applications. However, it is for this reason that we choose not to use it as our main test of non-Gaussian dependence, as it would not allow us to separately consider the marginal distributions and dependence structure of the likelihood.

For completeness, we show the distributions of additive transcovariance for the full-sky and cut-sky samples in \autoref{Fig:transcov}. There is a much larger excess of transcovariance in the cut-sky likelihood compared to the full-sky likelihood than is seen for the MI in \autoref{Fig:mi_all}. The fact that the transcovariance mixes the effects of non-Gaussian marginals and non-Gaussian dependence would prevent us from identifying whether this is due to the marginals, the dependence or both. As an additional check, we applied a probability integral transform to each pair such that the marginal distributions were Gaussian distributed, without affecting the dependence structure, and found that the resulting distribution had a much smaller cut-sky excess, similar to the MI in \autoref{Fig:mi_all}. 

\section{Non-Gaussian fields}
\label{Sec:nongauss_fields}

We have demonstrated in \autoref{Sec:fullsky} that a Gaussian likelihood is sufficient to obtain accurate parameter constraints in a combined weak lensing and galaxy clustering analysis on the full sky, and in \autoref{Sec:cut_sky} that the additional non-Gaussianity of the cut-sky likelihood is insufficient to introduce significant inaccuracy, both provided that the observable fields may be described using Gaussian statistics. We have reason to believe this to be a good approximation for the purposes of this study. 

First, the matter distribution is most Gaussian on linear scales, corresponding to low $\ell$, and most non-Gaussian at high $\ell$. But the inverse is true for the power spectrum likelihood: it is most Gaussian at high $\ell$, and most non-Gaussian at low $\ell$. While we have demonstrated this behaviour in the previous sections for Gaussian fields, we expect it to hold generally, as the number of $\alm$s contributing to each $\Cl$ estimate increases with $\ell$ regardless of the statistics of the field. 
The largest contribution to potential inaccuracy in the Gaussian likelihood therefore comes from linear scales, where the observable fields are well described as Gaussian.
Additionally, the presence of shape noise causes two further effects: it decreases the non-Gaussianity of the fields on all scales, and relatively upweights the contribution of large scales to the overall constraining power, as discussed in \autoref{Sec:robustness}. Both of these effects will increase the accuracy of the Gaussian fields assumption.
Finally, the process of going from galaxy catalogues to power estimates involves first averaging over galaxies in each pixel, followed by a spherical harmonic transform, both of which may be expected to approximately Gaussianise the $\alm$s following the Central Limit Theorem. We have tested the latter in simple tests using HEALPix spherical harmonic transforms of arbitrary non-Gaussian fields and found it to hold. Gaussian $\alm$s in turn imply approximately gamma-distributed auto-$\Cl$ estimates. However, the degrees of freedom in these gamma distributions may be reduced (and hence the non-Gaussianity increased) if the $\alm$s are correlated, similar to what happens on the cut sky. This idea was tested in \cite{Taylor2019}, which found no detectable difference between $\Cl$ distributions measured from Gaussian and lognormal simulations, which have been shown to well approximate real weak lensing data \citep{Taruya2002, Hilbert2011, Clerkin2016}. 

We note that non-Gaussian fields will introduce additional contributions to the $C_\ell$ covariance, which we have neglected here. Specifically, there is a contribution arising from
four-point correlations within a survey volume
-- often referred to as the connected non-Gaussian covariance --
and a generally larger contribution arising from
the dependence of such correlations on unmeasured super-survey modes
-- commonly termed super-sample covariance. \citep[See e.g.][for more details.]{Scoccimarro1999, Cooray2001, Takada2007, Takada2013, Li2014, Barreira2018, Barreira2018a}
These additional covariance contributions predominantly affect high $\ell$, which will have the effect of relatively upweighting the low-$\ell$ regime where the Gaussian likelihood is least accurate. However, we found in \autoref{Sec:robustness} that the Gaussian likelihood still performs well even at extremely low $\ell$, so we do not expect this to outweigh the other factors outlined above, and
we expect the overall impact of non-Gaussian fields on the accuracy of a Gaussian likelihood to be negligible. Nevertheless, here we apply the same techniques as in \autoref{Sec:cut_sky} to test the non-Gaussianity detected in a more realistic set of weak lensing simulations.

\subsection{Non-Gaussian fields: Simulations}

We use the SLICS\footnote{\href{https://slics.roe.ac.uk}{https://slics.roe.ac.uk}}, which are independent N-body weak lensing simulations, described in detail in \cite{Harnois-Deraps2015} and \cite{Harnois-Deraps2018}. We study the only available tomographic power spectra, which were the weak lensing convergence power spectra from the KiDS-450-like setup used in \cite{Hildebrandt2017}. These are flat-sky linearly-spaced bandpowers for auto-spectra only, produced from 948 independent realisations of $60\,\text{deg}^2$ sky patches. We use up to $\lmax = 5000$ in order to include non-linear scales for our tests.

We generated an equivalent batch of Gaussian-field simulations using pymaster, the Python implementation of NaMaster\footnote{\href{https://github.com/LSSTDESC/NaMaster}{https://github.com/LSSTDESC/NaMaster}} \citep{Alonso2019}, using a KiDS-450-like setup with four tomographic bins following the specification in Table 1 of \cite{Hildebrandt2017}.

Following the process in \autoref{Sec:cut_sky}, we now compare the non-Gaussianity of the marginals and dependence in the distributions from the SLICS compared to the Gaussian field sample.

\subsection{Non-Gaussian fields: Effect on marginal distributions}

\begin{figure}
\includegraphics[width=\columnwidth]{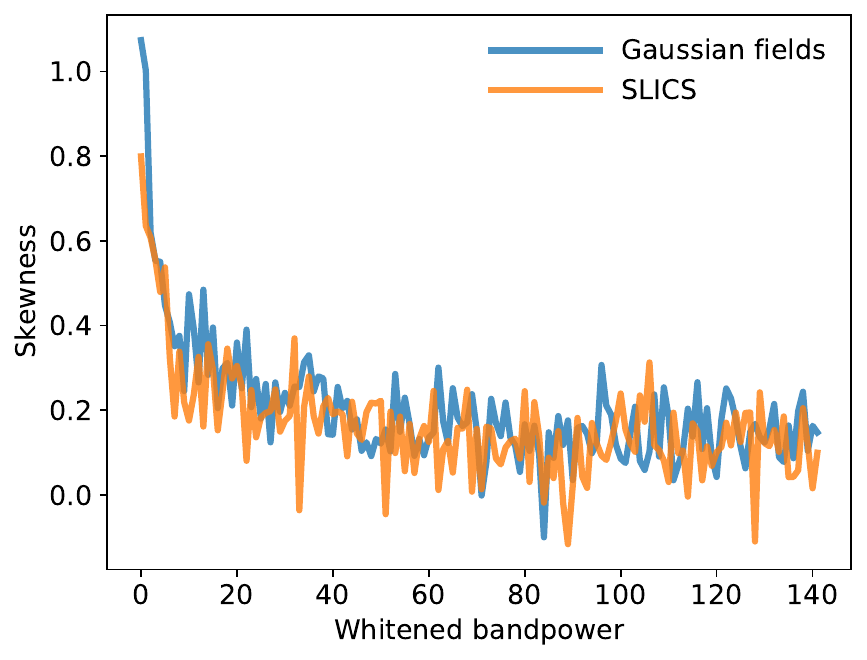}
\caption{Skewness of SLICS bandpowers compared to the Gaussian fields sample, after whitening to remove the effect of linear correlations.}
\label{Fig:slics_skew}
\end{figure}

\autoref{Fig:slics_skew} shows the skewness of the SLICS compared to the Gaussian field sample, averaged over the four redshift bins. We whitened each sample prior to calculating skewness, because we found that there were significant linear correlations present in the SLICS data that were not present in the Gaussian field simulations. These correlations are likely to be real rather than an artefact, since the SLICS were designed and validated specifically for covariance estimation \citep{Harnois-Deraps2015}. However, what matters for the accuracy of a multivariate Gaussian distribution is the non-Gaussianity of the marginals after whitening, since the Gaussian PDF effectively whitens the data vector itself. We find that after whitening, the skewness is consistent to within the level of the noise, with a possible slight excess for the Gaussian fields. We find similar consistency for the excess kurtosis. We conclude that there is no evidence of additional non-Gaussianity of the marginals for realistic non-Gaussian weak lensing fields.

\subsection{Non-Gaussian fields: Effect on dependence structure}

\begin{figure*}
\centering
\includegraphics[width=.95\textwidth]{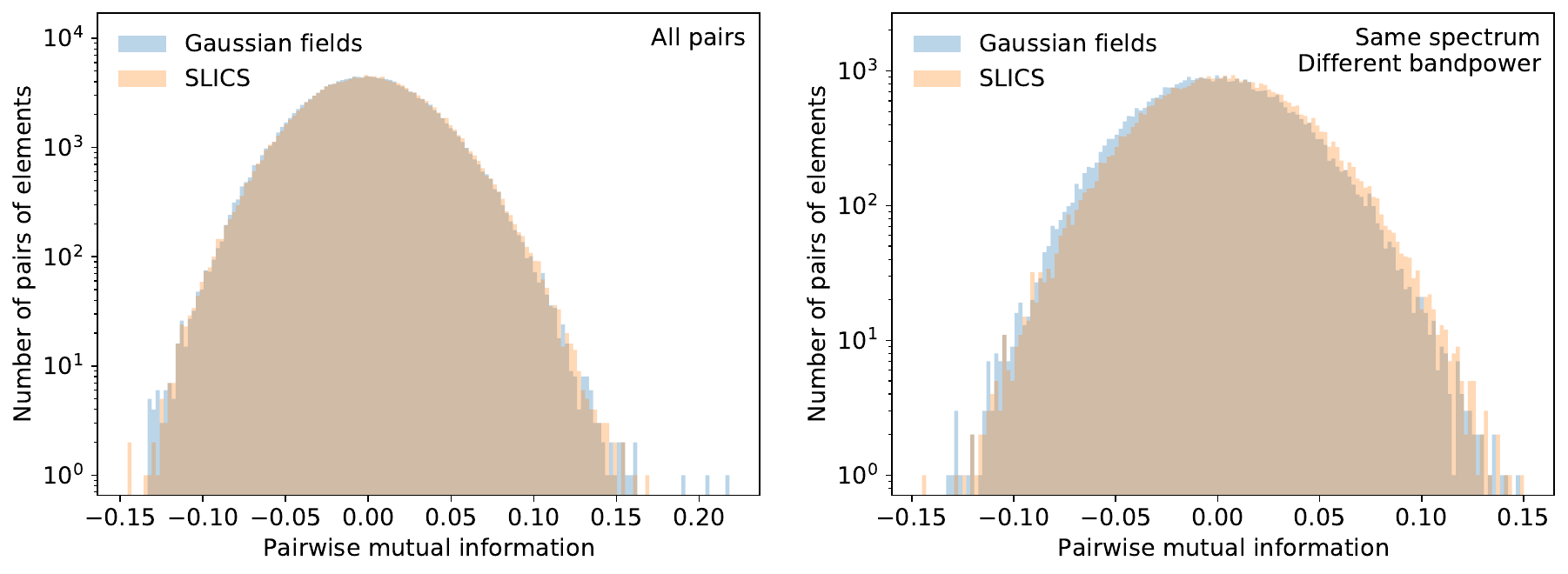}
\caption{Comparison of non-Gaussian dependence between the SLICS weak lensing simulations and a similar sample of Gaussian fields. Non-Gaussian dependence is quantified by pairwise mutual information after whitening. The left panel shows all pairs of data elements, while the right panel shows only those containing different bandpower pairs in the same spectrum. This population exhibits a possible small excess in non-Gaussian dependence.}
\label{Fig:slics_mi}
\end{figure*}

\begin{figure}
\includegraphics[width=\columnwidth]{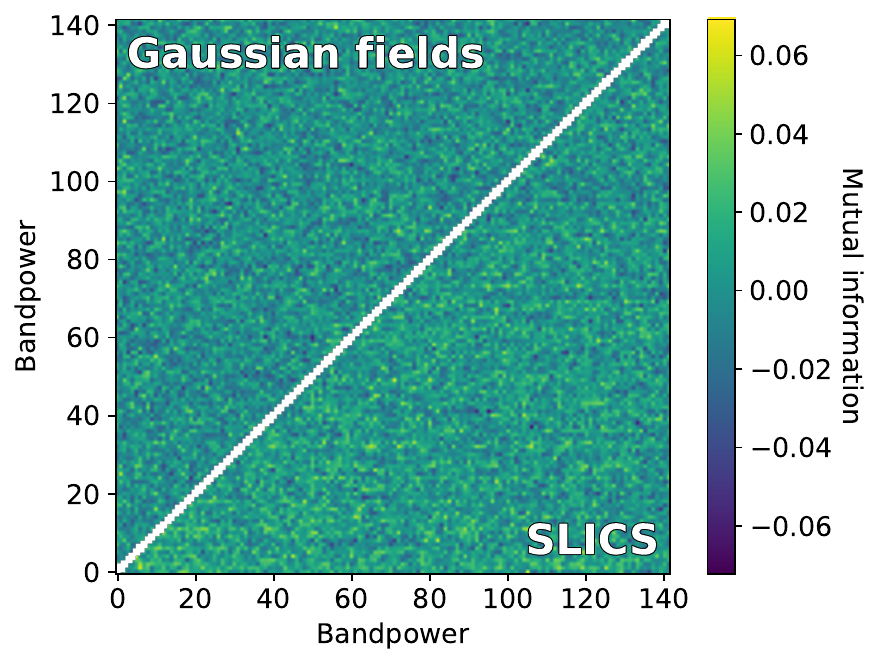}
\caption{Matrix of pairwise mutual information after pairwise whitening, averaged over four redshift bins, compared between the SLICS and the Gaussian fields sample.}
\label{Fig:slics_mi_matrix}
\end{figure}

Following the procedure described in \autoref{Sec:ma_dependence}, we measure pairwise mutual information (MI) after pairwise whitening and compare between the SLICS and the Gaussian field sample. We find that the overall MI distributions, shown in the left panel of \autoref{Fig:slics_mi}, are almost indistinguishable, but there is a very small excess for the SLICS. By splitting pairs of data elements into populations depending on their relationship, we find that this excess is due to a particular population: different bandpowers in the same spectrum, shown in the right panel of \autoref{Fig:slics_mi}. We find no apparent redshift dependence in this behaviour, nor does it have any apparent $\ell$-dependent structure: \autoref{Fig:slics_mi_matrix} shows the matrix of pairwise MI compared between the two samples. Both samples appear consistent with noise, with a slightly higher noise level in the SLICS. Whether this is a real or spurious effect is unknown; however, we can expect its effect on the accuracy of the Gaussian likelihood to be negligible: the average MI for the SLICS is $1.9 \times 10^{-4}$, far below the $9.0 \times 10^{-4}$ that we have shown to cause negligible inaccuracy in constraints obtained using the Gaussian likelihood in \autoref{Fig:leff_post}. Therefore, we find no evidence to suggest the conclusions drawn from our Gaussian field tests in \autoref{Sec:fullsky} and \autoref{Sec:cut_sky} should not hold for real weak lensing fields.
    
\section{Conclusions}
\label{Sec:conclusions}

It is well established that the true likelihood of weak lensing two-point statistics is non-Gaussian \citep{Sellentin2018, Sellentin2018a, DiazRivero2020, Louca2020}, and yet contemporary analyses routinely neglect this and assume a Gaussian likelihood \citep{Troxel2018, Hikage2019, Joachimi2020}. In this paper we have tested the impact of assuming a Gaussian likelihood for a \textit{Euclid}-like combined power spectrum analysis of weak lensing, galaxy clustering and their cross-correlation, on the inferred posterior distributions of dark energy parameters.

In \autoref{Sec:fullsky} we have found that on the full sky, the Gaussian likelihood returns the correct posterior maximum, two-dimensional contours and one-dimensional posterior probability density. This holds both when all other parameters are fixed or when marginalising over a third parameter, and for any choice of fiducial cosmology consistent with the data. The Gaussian likelihood is even a good approximation at low $\ell$, where the true likelihood is most non-Gaussian.

We have shown in \autoref{Sec:cut_sky} that a sky cut increases the non-Gaussianity of both the marginal distributions and dependence structure of the likelihood. However, by generating a mock cut-sky data vector and likelihood with the appropriate amount of non-Gaussianity in both cases, we have found that this additional non-Gaussianity introduces only negligible additional inaccuracy into the posterior parameter constraints obtained using the Gaussian likelihood.

The results presented in \autoref{Sec:fullsky} and \autoref{Sec:cut_sky} are obtained under the assumption of Gaussian fields. We have argued in \autoref{Sec:nongauss_fields} that this is a sufficient approximation for the purposes of this analysis. Nevertheless, we have compared results obtained under this approximation to those obtained using an equivalent set of N-body weak lensing simulations, and found no evidence of significant additional non-Gaussianity of the power spectrum likelihood.

Our results indicate that a Gaussian likelihood will be sufficient for robust cosmological inference with power spectra from stage IV weak lensing surveys such as \textit{Euclid}. This conclusion is further supported by the results obtained in \cite{Taylor2019}, which found no significant difference in parameter constraints obtained using a Gaussian likelihood compared to a likelihood-free approach. We cannot automatically extend this conclusion to the correlation function, which has a more complicated behaviour due to the mixing of scales \citep{Sellentin2018}. \cite{Lin2020} have found that a Gaussian likelihood is likely to be sufficiently accurate for parameter inference from LSST data. However, the disagreement between that result and that of \cite{Hartlap2009}, who found that the assumption of a Gaussian correlation function likelihood introduced significant inaccuracy in parameter constraints from a weak lensing analysis of the Chandra Deep Field South, remains to be fully understood.

\section*{Acknowledgements}

We are grateful to Joachim Harnois-D\'eraps for making available the SLICS data products, and to Alex Hall and Peter Taylor for conversations from which this project grew.
We also thank the anonymous referee for helpful and constructive feedback that improved the article.
REU acknowledges a studentship from the UK Science and Technology Facilities Council.
LW is supported by a UK Space Agency grant.
This work has made use of the following open-source software packages: CAMB \citep{Lewis2000, Howlett2012}, CosmoSIS \citep{Zuntz2015}, HEALPix/healpy \citep{Gorski2005, Zonca2019}, Matplotlib \citep{Hunter2007}, NaMaster \citep{Alonso2019}, NumPy \citep{Harris2020}, SciPy \citep{Virtanen2020}.

\section*{Data Availability}

The high-resolution simulated data used in \autoref{Sec:cut_sky} are made available at \href{https://dx.doi.org/10.5281/zenodo.4316733}{https://dx.doi.org/10.5281/zenodo.4316733}. The SLICS data used in \autoref{Sec:nongauss_fields} were provided by Joachim Harnois-D\'eraps by permission, and are available at \href{https://slics.roe.ac.uk}{https://slics.roe.ac.uk}. Other simulated data used in this article may be reproduced using scripts which will be shared on reasonable request to the corresponding author. 


\bibliographystyle{mnras}
\bibliography{Mendeley}






\bsp	
\label{lastpage}
\end{document}